\documentclass{ws-procs975x65}
\newcommand{\beq}{\begin{equation}}
\newcommand{\eeq}{\end{equation}}
\newcommand{\bdm}{\begin{displaymath}}
\newcommand{\edm}{\end{displaymath}}
\newcommand{\bctr}{\begin{center}}
\newcommand{\ectr}{\end{center}}

\newcommand{\om}{\Omega_M}
\newcommand{\oll}{\Omega_{\Lambda}}

\newcommand{\ob}{\Omega_{baryon}}
\newcommand{\obh}{\Omega_{baryon}h^{2}}
\newcommand{\oc}{\Omega_{CDM}}
\newcommand{\ok}{\Omega_{k}}
\newcommand{\og}{\Omega_{\gamma}}

\newcommand{\orel}{\Omega_{rel}}

\newcommand{\on}{\Omega_{\nu}}

\def\spose#1{\hbox to 0pt{#1\hss}}
\newcommand{\dg}           {\mbox{$^{\circ}$}}
\newcommand{\lsim}         {\mbox{$_<\atop^{\sim}$}}
\newcommand{\gsim}         {\mbox{$_>\atop^{\sim}$}}

\begin{document}

\title{
\raisebox{1.5ex}{\tiny{I}}
\raisebox{1.2ex}{\footnotesize{N}}
\raisebox{1.2ex}{\small{F}}
\raisebox{1.1ex}{\normalsize{L}}
\raisebox{0.9ex}{\large{A}}
\raisebox{0.7ex}{\Large{TI}}
\raisebox{0.2ex}{\huge{O}}
\Huge{N}\\
\small  and the \normalsize\\
\medskip
\raisebox{0.0ex}{C}
\raisebox{0.1ex}{o}
\raisebox{0.2ex}{s}
\raisebox{0.4ex}{m}
\raisebox{0.6ex}{i}
\raisebox{0.8ex}{c}
\raisebox{1.0ex}{ }
\raisebox{1.2ex}{M}
\raisebox{1.4ex}{i}
\raisebox{1.4ex}{c}
\raisebox{1.2ex}{r}
\raisebox{1.0ex}{o}
\raisebox{0.8ex}{w}
\raisebox{0.6ex}{a}
\raisebox{0.4ex}{v}
\raisebox{0.2ex}{e}
\raisebox{0.0ex}{ }
\raisebox{-0.2ex}{B}
\raisebox{-0.4ex}{a}
\raisebox{-0.6ex}{c}
\raisebox{-0.8ex}{k}
\raisebox{-0.6ex}{g}
\raisebox{-0.4ex}{r}
\raisebox{-0.2ex}{o}
\raisebox{-0.1ex}{u}
\raisebox{0.0ex}{n}
\raisebox{0.0ex}{d}}

\author{Charles H. Lineweaver}

\address{School of Physics, University of New South Wales,
Sydney, Australia\\
email: charley@bat.phys.unsw.edu.au}

\maketitle

\abstracts{I present a pedagogical review of inflation and  the cosmic
microwave background.
I describe how a short period of accelerated expansion can replace the special initial conditions of the
standard big bang model.
I also describe the development of CMBology: the study of the cosmic microwave background.
This cool (3 K) new cosmological tool is an increasingly precise rival and complement to many other methods in 
the race to determine the parameters of the Universe: its age, size, composition and detailed evolution.}

\section{A New Cosmology}

``The history of cosmology shows that in every age devout people believe that they have at last discovered
the true nature of the Universe.''\\
--  E. R. Harrison (1981)
\subsection{Progress}
Cosmology is the scientific attempt to answer fundamental questions of mythical proportion: 
How did the Universe come to be? How did it evolve? How will it end? 
If humanity goes extinct it will be of some solace to know that just before we went, incredible progress was 
made in our understanding of the Universe.
 ``The effort to understand the Universe is one of the very few things that lifts human life a 
little above the level of farce, and gives it some of the grace of tragedy.'' (Weinberg 1977).

A few decades ago cosmology was laughed at for being the only science with no data. 
Cosmology was theory-rich but data-poor.  It attracted armchair enthusiasts spouting 
speculations without data to test them. The night sky was calculated to be as bright as the 
Sun, the Universe was younger than the Galaxy and initial conditions, like animistic gods, 
were invoked to explain everything.
Times have changed. We have entered a new era of precision cosmology. 
Cosmologists are being flooded with high quality measurements from an army of new instruments.
We are observing the Universe at new frequencies, with higher sensitivity, higher 
spectral resolution and higher spatial resolution.  We have so much new data that 
state-of-the-art computers process and store them with difficulty.
Cosmology papers now include error bars -- often asymmetric and
sometimes even with a distinction made between statistical and systematic error bars. 
This is progress.

Cosmological observations such as measurements of the cosmic microwave background,
and the inflationary ideas used to interpret them, are at the heart of 
what we know about the origin of the Universe and everything in it. 
Over the past century cosmological observations have produced the standard hot big bang model
describing the evolution of the Universe in sharp mathematical detail. This model provides a 
consistent framework into which all relevant cosmological data seem to fit, and is the 
dominant paradigm against which all new ideas are tested.
It became the dominant paradigm in 1965 with the discovery of the cosmic microwave.
In the 1980's the big bang model was interpretationally upgraded to include an early 
short period of rapid expansion and a critical density of 
non-baryonic cold dark matter.

For the past 20 years many astronomers have assumed that 95\% of the Universe was
clumpy non-baryonic cold dark matter. They also assumed that the cosmological 
constant, $\oll$, was Einstein's biggest blunder and could be ignored.
However, recent measurements of the cosmic microwave background combined with supernovae and other
cosmological observations have given us a new inventory. We now find that $73\%$ of 
the Universe is made of vacuum energy, while only $23\%$ is 
made of non-baryonic cold dark matter. 
Normal baryonic matter, the stuff this paper is made of, makes up about
$4\%$ of the Universe.
Our new inventory has identified a previously unknown 73\% of the Universe! 
This has forced us to abandon the standard CDM ($\om = 1$) model and replace it with a new
hard-to-fathom $\Lambda$-dominated CDM model.

\subsection{Big Bang: Guilty of Not Having an Explanation}

``...the standard big bang theory says nothing about what banged, why it banged, or what happened before it banged.
The inflationary universe is a theory of the ``bang'' of the big bang.''  -- Alan Guth (1997). \\ %( 1997 p xiii).

Although the standard big bang model can explain much about the evolution of the Universe, there are
a few things it cannot explain:

\begin{itemize}
\item 
The Universe is clumpy.
Astronomers, stars, galaxies, clusters of galaxies and even larger structures are sprinkled about.
The standard big bang model cannot explain where this hierarchy of clumps came from-- it cannot explain the origin of
structure. We call this the structure problem. 

\item
In opposite sides of the sky, the most distant regions of the Universe are at almost
the same temperature. But in the standard big bang model they have never been in causal contact -- 
they are outside each other's causal horizons. Thus, the standard model cannot explain 
why such remote regions have the same temperature. We call this the horizon problem.

\item
As far as we can tell, the geometry of the Universe is flat -- the interior angles of large triangles add up
to $180 \dg$. If the Universe had started out with a tiny deviation from flatness, the standard big bang 
model would have quickly generated a measurable degree of non-flatness. The standard big bang model cannot 
explain why the Universe started out so flat.  We call this the flatness problem. 

\item
Distant galaxies are redshifted. The Universe is expanding. Why is it expanding? The standard big
bang model cannot explain the expansion. We call this the expansion problem.
\end{itemize}

Thus the big bang model is guilty of not having explanations for structure, homogeneous temperatures, flatness
or expansion.
It tries -- but its explanations are really only wimpy excuses called initial conditions.
These initial conditions are 
\begin{itemize}
\item the Universe started out with small seeds of structure
\item the Universe started out with the same temperature everywhere
\item the Universe started out with a perfectly flat geometry
\item the Universe started out expanding
\end{itemize} 

Until inflation was invented in the early 1980's, these initial conditions were tacked
onto the front end of the big bang. With these initial conditions, the evolution of the Universe
proceeds according to general relativity and can produce the Universe we see around us today.
Is there anything wrong with invoking these initial conditions? How else should the Universe have started?
The central question of cosmology is: How did the Universe get to be the way it is? Scientists have made a
niche in the world by not answering this question with ``That's just the way it is.'' And yet, that was the nature
of the explanations offered by the big bang model without inflation.

``The horizon problem is not a failure of the standard big bang theory in the strict sense, since it is neither an 
internal contradiction nor an inconsistency between observation and theory. The uniformity of the observed universe 
is built into the theory by postulating that the Universe began in a state of uniformity. As long as the uniformity 
is  present at the start, the evolution of the Universe will preserve it. The problem, instead, is one of predictive power. 
One of the most salient features of the observed universe -- its large scale uniformity -- cannot be explained by the 
standard big bang theory; instead it must be assumed as an initial condition.''\\
       -- Alan Guth (1997)   %p 184

The big bang model without inflation has special initial conditions tacked on to it in the first picosecond.
With inflation, the big bang doesn't need special initial conditions. It can do with inflationary expansion and
a new unspecial (and more remote) arbitrary set of initial conditions -- sometimes called chaotic initial conditions -- 
sometimes less articulately described as `anything'. The question that still haunts inflation 
(and science in general) is: Are arbitrary initial conditions a more realistic ansatz? 
Are theories that can use them as inputs more predictive?
Quantum cosmology seems to suggest that they are.
We discuss this issue more in Section \ref{sec:status}.

\section{Tunnel Vision: the Inflationary Solution}
\label{sec:expansion}

Inflation can be described simply as any period of the Universe's evolution in which the size of the Universe
is accelerating. This surprisingly simple type of expansion  leads to our observed universe without 
invoking special initial conditions.
The active ingredient of the inflationary remedy to the structure, horizon and flatness problems
is rapid exponential expansion sometime within the first picosecond ( = trillionth of a second = $10^{-12}$ s)
after the big bang.
If the structure, flatness and horizon problems are so easily solved,
it is important to understand how this quick cure works.
It is important to understand the details of expansion and cosmic horizons. 
Also, since our Universe is becoming more $\Lambda$-dominated every day (Fig. \ref{fig:omegai}), 
we need to prepare for the future. Our descendants will, of necessity,  become more and more 
familiar with inflation, whether they like it or not. Our Universe is surrounded by inflation
at both ends of time. 

\subsection{Friedmann-Robertson-Walker metric $\rightarrow$ Hubble's law and Cosmic Event Horizons}
The general relativistic description of an homogeneous, isotropic universe is based upon the 
Friedmann-Robertson-Walker (FRW) metric for which the spacetime interval $ds$, between two events, is given by
\beq  
ds^2 = -c^2dt^2 + R(t)^2[d\chi^2+S_k^2(\chi)d\psi^2], 
\label{eq:FRW}
\eeq
where $c$ is the speed of light, $dt$ is the time separation, $d\chi$ is the comoving coordinate separation 
and $d\psi^2=d\theta^2+sin^2\theta d\phi^2$, where $\theta$ and $\phi$ are the polar and azimuthal angles in 
spherical coordinates. The scale factor $R$ has dimensions of distance. 
The function $S_k(\chi)=\sin\chi$, $\chi$ or $\sinh\chi$ for closed (positive $k$), flat ($k=0$) or 
open (negative $k$) universes respectively (see e.g. Peacock 1999 p. 69).
%A 3-D surface of a 4-sphere is a closed geometry. 
%A 3-D surface of a 4-table is a flat geometry and the
%3-D surface of a 4-saddle is an open geometry.

\begin{figure}[t!h]
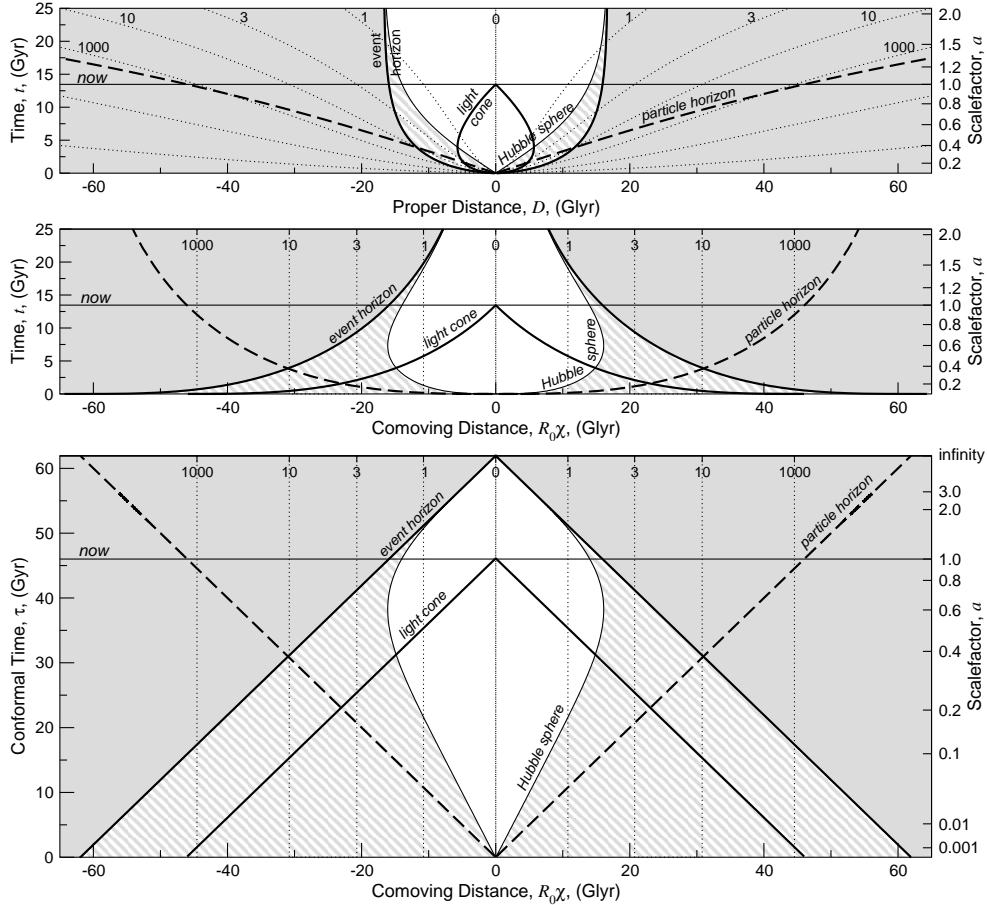

%\figurebox{22pc}{15pc}{} % to have a box alone
%%%\epsfxsize=32pc % will enlarge or reduce the postscript figures based on the xsize
%%%\epsfbox{fig1triptych.ps} % postscript image file name
\psfig{file=dist.eps,width=130mm}
\psfig{file=comov.eps,width=130mm}
\psfig{file=conf.eps,width=130mm}
\caption{{\bf Expansion of the Universe.}   %\\
We live on the central vertical worldline.
The dotted lines are the worldlines of galaxies being expanded away from us as 
the Universe expands. They are labeled by the redshift of their light that is reaching us 
today, at the apex of our past light cone. 
Top: 
In the immediate past our past light cone is shaped like a cone. But as we follow it further 
into the past it curves in and makes a teardrop shape. This is a fundamental 
feature of the expanding universe; the furthest light that we can see now was receding from us 
for the first few billion years of its voyage.  The Hubble sphere, particle horizon, event horizon
and past light cone are also shown  (Eqs.~\ref{eq:hs} -- \ref{eq:lc}).
Middle:
We remove the expansion of the Universe from the top panel by plotting comoving distance 
on the x axis rather than proper distance. Our teardrop-shaped light cone then becomes a 
flattened cone and the constant proper distance of the event horizon becomes a shrinking 
comoving event horizon -- the active ingredient of inflation (Section \ref{sec:magic}). 
Bottom:  the radius of the current observable Universe (the particle horizon) 
is 47 billion light years (Glyr), i.e., the most distant galaxies that we can see on our past light 
cone are now 47 billion light years away. 
The top panel is long and skinny because the Universe is that way -- the Universe is larger 
than it is old -- the particle horizon is 47 Glyr while the age is only 13.5 Gyr -- thus  
producing the $3:1 (\approx 47:13.5)$ aspect ratio. In the bottom panel, space and time are on the 
same footing in conformal/comoving coordinates and this produces the $1:1$ aspect ratio. 
For details see Davis \& Lineweaver (2003).
}
\label{fig:triptych}
\end{figure}

In an expanding universe, the proper distance $D$ between an observer at the origin and a 
distant galaxy is defined to be along a surface of constant time ($dt=0$).  
We are interested in the radial distance so $d\psi=0$. The FRW metric then reduces to $ds = R d\chi$ which, 
upon integration, becomes,
\beq  
D(t) = R(t)\chi.
\eeq
Taking the time derivative and assuming that we are dealing with a comoving 
galaxy ($\dot{\chi} = 0$) we have,
\bea
v(t) &=& \dot{R}(t) \chi,\\
v(t) &=& \frac{\dot{R}(t)}{R}\; R \;\chi,\\
{\mbox \rm Hubble's \ Law} \;\;\;v(t) &=& H(t) D,\\
{\mbox \rm Hubble \ Sphere} \;\;\;D_{H} &=& c/H(t).
\label{eq:hs}
\eea
The Hubble sphere is the distance at which the recession velocity $v$ is equal to the speed of light. 
Photons have a peculiar velocity of $c = \dot{\chi} R$, or equivalently photons move through comoving space
with a velocity $\dot{\chi} = c/R$. The comoving distance traveled by a photon is
$\chi = \int \dot{\chi} dt$, which we can use to define the comoving coordinates of some fundamental concepts:
\beq
{\mbox \rm Particle \ Horizon} \;\;\; \chi_{ph}(t) = c\; \int_{0}^{t} dt/R(t),
\label{eq:ph}
\eeq 
\beq
{\mbox \rm Event \ Horizon} \;\;\; \chi_{eh}(t) = c\; \int_{t}^{\infty} dt/R(t),
\label{eq:eh}
\eeq 
\beq
{\mbox \rm Past \ Light \ Cone} \;\;\; \chi_{lc}(t) = c\; \int_{t}^{t_{o}} dt/R(t).
\label{eq:lc}
\eeq 
Only the limits of the integrals are different.
The horizons, cones and spheres of Eqs. \ref{eq:hs} - \ref{eq:lc}
are plotted in Fig. \ref{fig:triptych}.

%\clearpage
%%%%%%%%%%%%%%%%%%%%%%%%%%%%%%%%%%%%%%%%%%%%%%%%%%%%%%%%%%%%%%%%%%%%%%%%%%%%%%%%%%%%%%%%%%%%%%%%%%%%%%%%%%%%%%%%%%%%%%%
%%%%%%%%%%%%%%%%%%%%%%%%%%%%%%%%%%%%%%%%%%%%%%%%%%%%%%%%%%%%%%%%%%%%%%%%%%%%%%%%%%%%%%%%%%%%%%%%%%%%%%%%%%%%%%%%%%%%%%%
%

%%%%%%%%%%%%%%%%%%%%%%%%%%%%%%%%%%%%%%%%%%%%%%%%%%%%%%%%%%%%%%%%%%%%%%%%%%%%%%%%%%%%%%%%%%%%%%%%%%%%%%%%%%%%%%%%%%%%%%%
\begin{figure}[!h]
%\figurebox{22pc}{15pc}{} % to have a box alone
\epsfxsize=33pc % will enlarge or reduce the postscript figures based on the xsize
\epsfbox{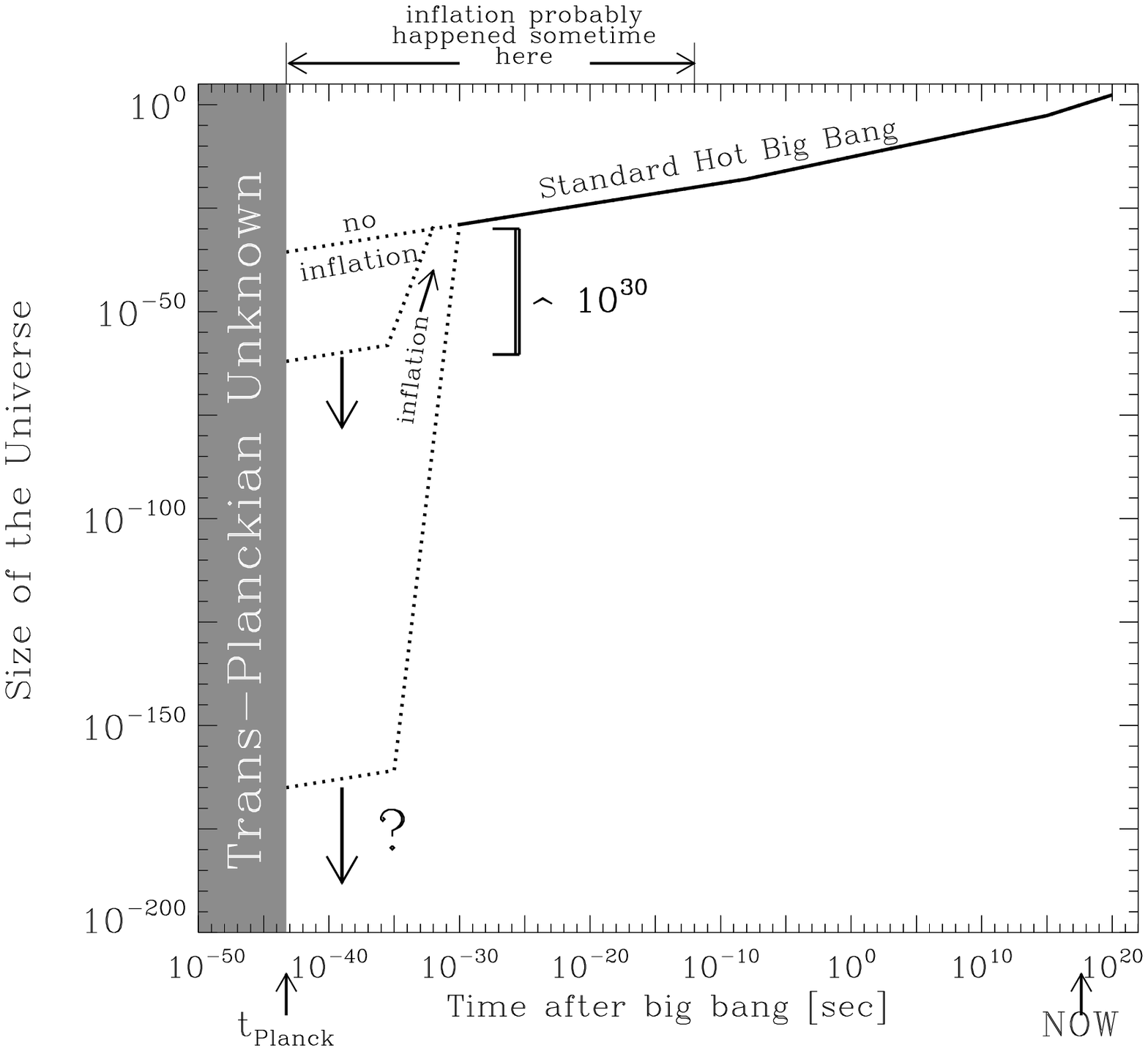} % postscript image file name
\caption{ 
Inflation is a short period of accelerated expansion that probably happened sometime within the first
picosecond ($10^{-12}$ seconds) -- during which the size of the Universe grows by more than a factor of
$\sim 10^{30}$. 
The size of the Universe coming out of the `Trans-Planckian Unknown' is unknown. Compared to its size today,
maybe it was $10^{-60}$ as shown in one model\dots or maybe it was $10^{-165}$ as shown in the other model\dots
or maybe even smaller (hence the question mark).
In the two models shown, inflation starts near the GUT scale, ($\sim 10^{16}$ GeV or $\sim 10^{-35}$ seconds)
and ends at about $10^{-30}$ seconds after the bang.}
\label{fig:Inflationaryexpansion}
\end{figure}
%%%%%%%%%%%%%%%%%%%%%%%%%%%%%%%%%%%%%%%%%%%%%%%%%%%%%%%%%%%%%%%%%%%%%%%%%%%%%%%%%%%%%%%%%%%%%%%%%%%%%%%%%%%%%%%%%%%%%%%

\subsection{Inflationary Expansion: The Magic of a Shrinking Comoving Event Horizon}
\label{sec:magic}

Inflation doesn't make the observable universe big. The observable universe is as big as it is. 
What inflation does is make the region from which the Universe emerged, very small. 
How small? is unknown (hence the question mark in Fig.~\ref{fig:Inflationaryexpansion}), but
small enough to allow the points in opposite sides of the sky ({\bf A} and {\bf B} in Fig.~\ref{fig:fig1bot})
to be in causal contact.

The exponential expansion of inflation produces an event horizon at a constant proper distance which is equivalent to
a shrinking comoving horizon. 
A shrinking comoving horizon is the key to the inflationary solutions of the 
structure, horizon and flatness problems. So let's look at these concepts carefully in 
Fig.~\ref{fig:triptych}.

The new $\Lambda$-CDM cosmology has an event horizon and it is this cosmology that is plotted
in Fig.~\ref{fig:triptych} (the old standard CDM cosmology did not have an event horizon).
To have an event horizon means that there will be events in the Universe that we will never be able
to see no matter how long we wait. This is equivalent to the statement that the expansion of the 
Universe is so fast that it prevents some distant light rays, that are propagating toward us, 
from ever reaching us.
In the top panel, one can see the rapid expansion of objects away from the central observer.
As time goes by, $\Lambda$ dominates and
the event horizon approaches a constant physical distance from an observer. Galaxies do not
remain at constant distances in an expanding universe.  
Therefore distant galaxies keep leaving the horizon, i.e., with time, they move upward and 
outward along the lines labeled with redshift `1' or `3' or `10'. 
As time passes,
fewer and fewer objects are left within the event horizon. The ones that are left, started out very close to 
the central observer.  Mathematically, the $R(t)$ in the denominator of Eq.~\ref{eq:eh} increases so fast
that the integral converges. As time goes by, the lower limit $t$ of the integral gets bigger, making
the integral converge on a smaller number -- hence the comoving event horizon shrinks.
The middle panel shows clearly that in the future, as $\Lambda$ increasingly dominates the dynamics of the Universe, the
comoving event horizon will shrink. This shrinkage is happening slowly now but during inflation
it happened quickly.
The shrinking comoving horizon in the middle panel of Fig.~\ref{fig:triptych} is a slow
and drawn out version of what happened during inflation -- so we can use what is going on now to 
understand how inflation worked in the early universe.
In the middle panel galaxies move on vertical lines upward, while
the comoving event horizon shrinks. As time goes by we are able to see a smaller and smaller region of
comoving space. Like using a zoom lens, or doing a PhD, we are able to see only a tiny
patch of the Universe, but in amazing detail.
Inflation gives us tunnel vision. The middle panel shows the narrowing of the tunnel.
Galaxies move up vertically and like objects falling into black holes, from our point of view
they are redshifted out of existence.

The bottom line is that accelerated expansion produces an event horizon at a given physical size and that
any particular size scale, including quantum scales, expands with the Universe and quickly becomes larger than 
the given physical size of the event horizon.

%%%%%%%%%%%%%%%%%%%%%%%%%%%%%%%%%%%%%%%%%%%%%%%%%%%%%%%%%%%%%%%%%%%%%%%%%%%%%%%%%%%%%%%%%%%%%%%%%%%%%%%%%%%%%%%%%%%%%%%
\begin{figure}[t!h]
%\figurebox{22pc}{15pc}{} % to have a box alone
\epsfxsize=33pc % will enlarge or reduce the postscript figures based on the xsize
\epsfbox{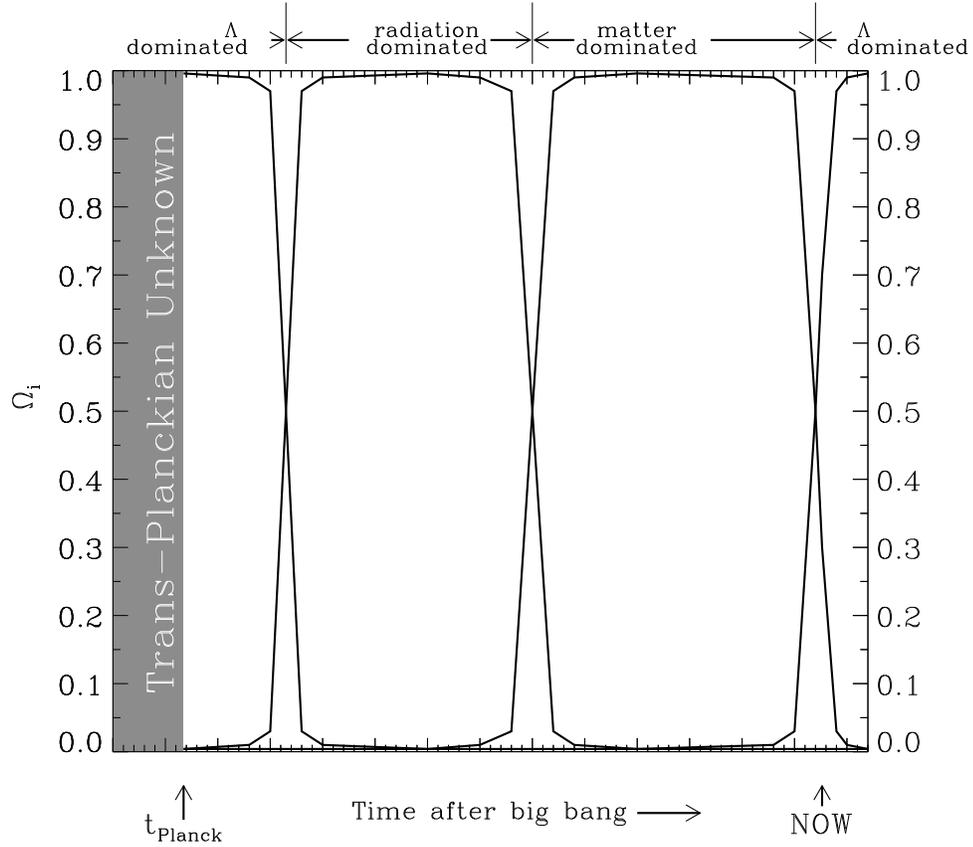} % postscript image file name
%\epsfbox{Omegaoft.ps} % postscript image file name
\caption{{\bf Friedmann Oscillations: The rise and fall of the dominant components of the Universe.}
The inflationary period can be described by a universe dominated by a large cosmological constant 
(energy density of a scalar field).
During inflation and reheating the potential of the scalar field is turned into massive particles which quickly decay into
relativistic particles and the Universe becomes radiation-dominated. Since $\rho_{rel} \propto R^{-4}$ and
$\rho_{matter} \propto R^{-3}$, as the Universe expands a radiation-dominated epoch gives way to
a matter-dominated epoch at $z \approx 3230$. And then, since $\rho_{\Lambda} \propto R^{o}$, the matter-dominated epoch
gives way to a $\Lambda$-dominated epoch at $z \approx 0.5$.
Why the initial $\Lambda$-dominated epoch became a radiation-dominated epoch is not as easy to understand
as these subsequent oscillations governed by the Friedmann Equation (Eq.~\ref{eq:fried1}).
Given the current values $(h,\Omega_{m},\Omega_{\Lambda}, \Omega_{rel}) = (0.72, 0.27, 0.73, 0.0)$ 
the Friedmann Equation enables us to trace back through time the oscillations in the quantities
$\Omega_{m},\Omega_{\Lambda}$ and $\Omega_{rel}$.
}
\label{fig:omegai}
\end{figure}
%%%%%%%%%%%%%%%%%%%%%%%%%%%%%%%%%%%%%%%%%%%%%%%%%%%%%%%%%%%%%%%%%%%%%%%%%%%%%%%%%%%%%%%%%%%%%%%%%%%%%%%%%%%%%%%%%%%%%%%

\section{Friedmann Oscillations: The Rise and Fall of Dominant Components.}
Friedmann's Equation can be derived from Einstein's 4x4 matrix equation of general relativity
(see for example Landau \& Lifshitz 1975, Kolb and Turner 1992 or Liddle \& Lyth 2000):
% LLeq. 2.12).
%
\be
 R_{\mu \nu} - \frac{1}{2}g_{\mu \nu} {\mathcal R} = 8\pi G\; T_{\mu \nu} + \Lambda g_{\mu \nu}
\label{eq:GR}
\ee
where $ R_{\mu \nu}$ is the Ricci tensor, ${\mathcal R}$ is the Ricci scalar, $g_{\mu \nu}$ is the metric tensor
describing the local curvature of space (intervals of spacetime are described by $ds^2 =g_{\mu \nu}dx^{\mu}dx^{\nu}$), 
$T_{\mu \nu}$ is the stress-energy tensor and $\Lambda$ is the cosmological constant.
Taking the $(\mu, \nu) = (0,0)$ terms of Eq.~\ref{eq:GR}
and making the identifications of the metric tensor with the terms in
the FRW metric of Eq.~\ref{eq:FRW}, yields the Friedmann Equation:
\beq
H^{2}    =  \frac{8\pi G \rho}{3} - \frac{k}{R^{2}} + \frac{\Lambda}{3}               \label{eq:fried1}\\
\eeq
where $R$ is the scale factor of the Universe, $H = \dot{R}/R$ is Hubble's constant, 
$\rho$ is the density of the Universe in relativistic or non-relativistic matter, $k$ is 
the constant from Eq.~\ref{eq:FRW} and $\Lambda$ is the cosmological constant.
In words: the expansion ($H$) is controlled by the density ($\rho$), the geometry ($k$) and the cosmological
constant ($\Lambda$).
Dividing through by $H^{2}$ yields
\beq
1 = \frac{\rho}{\rho_{c}} - \frac{k}{H^{2}R^{2}} + \frac{\Lambda}{3H^{2}}             \label{eq:fried2}
\eeq
where the critical density $\rho_{c} = \frac{3 H^{2}}{8 \pi G}$.
Defining $\Omega_{\rho} = \frac{\rho}{\rho_{c}}$ and $\Omega_{\Lambda} = \frac{\Lambda}{3H^{2}}$
and using $\Omega = \Omega_{\rho} + \Omega_{\Lambda}$ we get,
\beq
1 - \Omega  = \frac{-k}{H^{2}R^{2}}                                                     \label{eq:fried3}
\eeq
or equivalently,
\beq
(1 - \Omega)H^{2}R^{2}  = constant.                                                      \label{eq:fried4}
\eeq
If we are interested in only post-inflationary expansion in the radiation- or matter-dominated epochs
we can ignore the $\Lambda$ term and multiply Eq.~\ref{eq:fried1} by $\frac{3}{8\pi G\rho}$ to get
\beq
\frac{3H^{2}}{8\pi G \rho}    =  1 - \frac{3k}{8\pi G \rho R^{2}}               \label{eq:fried5}
\eeq
which can be rearranged to give
\beq
\left(\Omega^{-1} - 1\right)\rho R^{2}    =  constant                      \label{eq:fried6}
\eeq

A more heuristic Newtonian analysis can also be used to derive Eqs.~\ref{eq:fried4} \& \ref{eq:fried6}
(e.g. Wright 2003).
Consider a spherical shell of radius $R$ expanding at a velocity $v = HR$, in a universe of density $\rho$.
Energy conservation requires, 
\beq
2E = v^{2} - \frac{2GM}{R}= H^{2}R^{2} - \frac{8\pi G R^{2}\rho}{3}. 
\label{eq:econservation}
\eeq
By setting the total energy equal to zero we obtain a critical density at which
$v=HR$ is the escape velocity,
\beq
\rho_{c} = \frac{3 H^{2}}{8 \pi G} = 1.879 \;h^{2} \times 10^{-29} g\; cm^{-3} \sim  20\; protons\; m^{-3}.
\label{eq:rhocrit}
\eeq
However, by requiring only energy conservation ($2E=constant$ not necessarily $E=0$) in Eq.~\ref{eq:econservation}, 
we find,
\beq
constant = H^{2}R^{2} - \frac{8\pi G R^{2}\rho}{3}.
\label{eq:econservation2}
\eeq
Dividing Eq.~\ref{eq:econservation2} by $H^{2}R^{2}$ we get
\beq
(1 - \Omega)H^{2}R^{2} = constant,
\label{eq:dyn1}
\eeq
which is the same as Eq.~\ref{eq:fried4}.
Multiplying Eq.~\ref{eq:econservation2} by $\frac{3}{8\pi G\rho R^{2}}$ we get
\beq
(\Omega^{-1} - 1)\rho R^{2} = constant 
\label{eq:dyn2}
\eeq
which is the same as Eq.~\ref{eq:fried6}.

\subsection{Friedmann's Equation $\rightarrow$ Exponential Expansion}
One way to describe inflation is that during inflation, a $\Lambda_{\rm inf}$ term dominates Eq.~\ref{eq:fried1}.
Thus, during inflation we have,
\bea
H^{2}                                  &=&     \frac{\Lambda_{\rm inf}}{3}  \label{eq:lamb}\\
\frac{dR}{dt}                          &=&     R\sqrt{\frac{\Lambda_{\rm inf}}{3}}\\
%\frac{dR}{R}                           &=&     dt\sqrt{\frac{\Lambda_{inf}}{3}}\\
\int^{R}_{R_{i}}\frac{dR}{R}   &=&     \int^{t}_{t_{i}}dt\sqrt{\frac{\Lambda_{\rm inf}}{3}}\\
ln \frac{R}{R_{i}}  &=&  \sqrt{\frac{\Lambda_{\rm inf}}{3}}\;(t - t_{i})\\
R             &\approx & R_{i} e^{Ht} \label{eq:last}
\eea
where $t_{i}$ and $R_{i}$ are the time and scale factor at the beginning of inflation. 
To get Eq.~\ref{eq:last} we have assumed $0 \approx t_{i} << t < t_{e}$
(where $t_{e}$ is the end of inflation) and we 
have used Eq.~\ref{eq:lamb}.
Equation \ref{eq:last} is the exponential expansion of the Universe during inflation.
The e-folding time is $1/H$. The doubling time is $(ln 2)/H$. That is,
during every interval $\Delta t = 1/H$, the size of the Universe increases by a factor of $e=2.718281828...$
and during every interval $\Delta t = (ln 2)/H$ the size of the Universe doubles.

\section{Inflationary Solutions to the Flatness and Horizon Problems}
\label{sec:horizon}

\subsection{What is the Flatness Problem?}
First I will describe the flatness problem and then the inflationary solution to it.
Recent measurements of the total density of the Universe find $ 0.95 < \Omega_{o} < 1.05$ (e.g. Table 1).
This near flatness is a problem because the Friedmann Equation tells us that $\Omega \sim 1$ is
a very unstable condition -- like a pencil balancing on its point. It is a very special condition that
won't stay there long. Here is an example of how special it is.
Equation~\ref{eq:fried6} shows us that  $(\Omega^{-1} - 1)\rho R^{2} = constant$.
Therefore, we can write,
\beq
(\Omega^{-1} - 1)\rho R^{2}     =    (\Omega_{o}^{-1} - 1)\rho_{o} R_{o}^{2}
\eeq
where the right hand side is today and the left hand side is at any arbitrary time.
We then have,
\beq
(\Omega^{-1} - 1)               =    (\Omega_{o}^{-1} - 1)\frac{\rho_{o}}{\rho} \left(\frac{R_{o}}{R}\right)^{2}.
\eeq
Redshift is related to the scale factor by $R = R_{o}/(1+z)$.
Consider the evolution during matter-domination where $\rho = \rho_{o}(1+z)^{3}$. 
%During radiation domination we have $\rho = \rho_{o}(1+z)^{4}$ 
Inserting these we get,
%matter domination
\beq
(\Omega^{-1} - 1)               =    \frac{(\Omega_{o}^{-1} - 1)}{1+z}.
\eeq
%radiation domination
%\beq
%(\Omega^{-1} - 1)               =    \frac{(\Omega_{o}^{-1} - 1)}{(1+z)^{2}}
%\eeq
Inserting the current limits on the density of the Universe, $ 0.95 < \Omega_{o} < 1.05$ (for which
$  -0.05  < (\Omega_{o}^{-1} - 1) < 0.05$), we get a constraint on the possible values that  $\Omega$ could have
had at redshift $z$,
\beq
\frac{1}{1+\frac{0.05}{1+z}} < \Omega < \frac{1}{1-\frac{0.05}{1+z}}.      \label{eq:z}               
\eeq
At recombination (when the first hydrogen atoms were formed) $z \approx 10^{3}$ and the constraint on $\Omega$ yields,
\beq
0.99995 < \Omega < 1.00005 
\eeq
So the observation that $0.95 < \Omega_{o} < 1.05$ today, means that at a redshift of $z \sim 10^{3}$ we must have had
$0.99995 < \Omega < 1.000005$.
This range is small...special.
However, $\Omega$ had to be even more special earlier on.
We know that the standard big bang successfully predicts the relative abundances of the
light nuclei during nucleosynthesis between $\sim 1$ minute and $\sim 3$ minutes after the 
big bang, so let's consider the slightly earlier time, 1 second after the big bang which is 
about the beginning of the epoch in which we are confident that the Friedmann Equation holds. 
The redshift was $z \sim 10^{11}$ and the resulting constraint on the density at that time was,
\beq
0.9999999999995 < \Omega < 1.0000000000005 
\eeq
This range is even smaller and more special,
(although I have assumed matter domination for this calculation, at redshifts higher than $z_{eq} \sim 3000$, we have
radiation domination and $\rho = \rho_{o}(1+z)^{4}$.  This makes the $1+z$ in Eq.~\ref{eq:z} a $(1+z)^{2}$ and
requires that early values of $\Omega$ be even closer to $1$ than calculated here).

To summarize:
\bea
                     0.95 &< \Omega_{o}(z=0)        <& 1.05                               \\
                   0.99995 &< \Omega(z=10^{3})   <& 1.000005                          \\
           0.9999999999995 &< \Omega(z=10^{11})  <& 1.0000000000005 
\eea
If the Friedmann Equation is valid at even higher redshifts, $\Omega$ must have been
even closer to one. 
These limits are the mathematical quantification behind our previous statement that:
`If the Universe had started out with a tiny deviation from flatness, the standard big bang 
model would have quickly generated a measurable degree of non-flatness.'
{\it If} we assume that $\Omega$ could have started out with any value, then we have a compelling question:
Why should $\Omega$ have been so fine-tuned to $1$?

Observing $\Omega_{o} \approx 1$ today can be compared to a pencil standing on its point. If you walk into a room and 
find a pencil standing on its point you think: pencils don't usually stand on their points. If a pencil is that way 
then some mechanism must have recently set it up because pencils won't stay that way long. 
Similarly, if you wake up in a universe
that you know would quickly evolve away from $\Omega = 1$ and yet you find that $\Omega_{o}=1$ then some mechanism must have
balanced it very exactly at $\Omega = 1$.

Another way to state this flatness problem is as an oldness problem. If $\Omega_{o} \approx 1$ today, then the Universe
cannot have gone through many e-folds of expansion which would have driven it away from $\Omega_{o} = 1$.
It cannot be very old. If the pencil is standing on its end, then
the mechanism to push it up must have just finished.
But we see that the Universe {\it is} old in the sense that it {\it has} gone through many e-foldings of expansion
(even without inflation).

If early values of $\Omega$ had exceeded $1$ by a tiny amount
then this closed Universe would have recollapsed on itself almost immediately. How did the Universe get to be so old?
If early values of $\Omega$ were less than $1$ by a tiny amount
then this open Universe would have expanded so quickly that no stars or galaxies would have formed. How did our
galaxy get to be so old?
The tiniest deviation from $\Omega = 1$ grows quickly into a collapsing
universe or one that expands so quickly that clumps have no time to form.

\subsection{Solving the Flatness  Problem}
How does inflation solve this flatness problem?
How does inflation set up the condition of $\Omega = 1$?
Consider Eq.~\ref{eq:fried4}: $(1 - \Omega)H^{2}R^{2}  = constant$.
During inflation $H = \sqrt{\Lambda_{\rm inf}/3} = constant$ (Eq.~\ref{eq:lamb}) and
 the scale factor $R$ increases by many orders of magnitude, $\gsim 10^{30}$. 
One can then see from Eq.~\ref{eq:fried4} that the large increase in the 
scale factor $R$ during inflation, with $H$ constant,
drives $\Omega \rightarrow 1$.
This is what is meant when we say that inflation
makes the Universe spatially flat.
In a vacuum-dominated expanding universe, $\Omega =1$ is a stable fixed point.
During inflation $H$ is constant and $R$ increases exponentially.
Thus, no matter how far $\Omega$ is from 1 before inflation, the exponential increase of $R$ during inflation
quickly drives it to $1$ and this is equivalent to flattening the Universe. 
Once driven to $\Omega = 1$ by inflation, the Universe will naturally evolve away from $\Omega = 1$ in the absence
of inflation as we showed in the previous section.

%\clearpage
%%%%%%%%%%%%%%%%%%%%%%%%%%%%%%%%%%%%%%%%%%%%%%%%%%%%%%%%%%%%%%%%%%%%%%%%%%%%%%%%%%%%%%%%%%%%%%%%%%%%%%%%%%%%%%%%%%%%%%%
\begin{figure}[!th]
%\figurebox{22pc}{15pc}{} % to have a box alone
\epsfxsize=30pc % will enlarge or reduce the postscript figures based on the xsize
\epsfbox{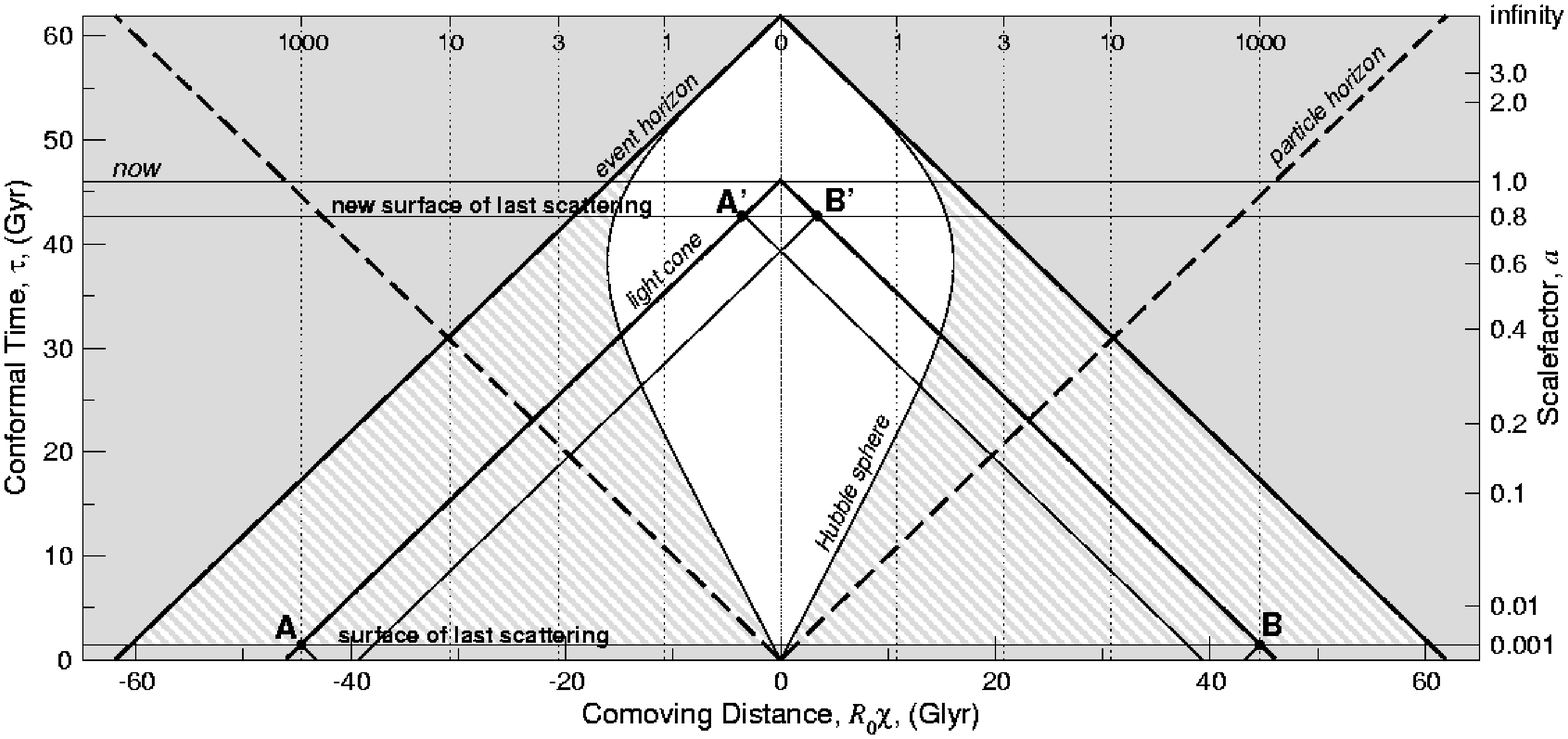} % postscript image file name
%\epsfbox{fig1bot1new.eps} % postscript image file name
\caption{Inflation shifts the position of the surface of last scattering.   %\\
Here we have modified the lower panel of Fig.~\ref{fig:triptych} to show what the insertion of an early
period of inflation does to the past light cones of two points, {\bf A} and {\bf B}, at the
surface of last scattering on opposite sides of the sky. 
An opaque wall of electrons -- the cosmic photosphere, also known as the surface of last scattering  -- 
is at a scale factor $a = R/R_{o} \approx 0.001$ when the Universe was $\approx 1000$ times smaller than it is now and 
only $380,000$ years old. 
The past light cones of {\bf A} and {\bf B} do not overlap -- they have never seen each other 
-- they have never been in causal contact. 
And yet we observe these points to be at the same temperature. This is the horizon problem (Sect. \ref{sec:horizonprob}).
Grafting an early epoch of inflation onto the big bang model moves the surface of last scattering upward to the line 
labeled ``new surface of last scattering''.  Points {\bf A} and {\bf B} move upward to {\bf A}$^{\prime}$ 
and {\bf B}$^{\prime}$. Their new past light cones overlap substantially. 
They have been in causal contact for a long time. 
Without inflation there is no overlap. With inflation there is.
That is how inflation solves the problem of identical temperatures in `different' horizons.
 The y axis shows all of time. That is, the range in conformal time $[0,62]$ Gyr corresponds
to the cosmic time range $[0, \infty]$ (conformal time $\tau$ is defined by $d\tau = dt/R$). 
Consequently, there is an upper limit to the size of the observable universe. 
The isosceles triangle of events within the event horizon are the only events in the Universe
that we will ever be able to see  -- probably a very small fraction of the entire universe.
That is, the x axis may extend arbitrarily far in both directions. Like this $\downarrow$.
}
\label{fig:fig1bot}
\end{figure}
%%%%%%%%%%%%%%%%%%%%%%%%%%%%%%%%%%%%%%%%%%%%%%%%%%%%%%%%%%%%%%%%%%%%%%%%%%%%%%%%%%%%%%%%%%%%%%%%%%%%%%%%%%%%%%%%%%%%%%%
%%%%%%%%%%%%%%%%%%%%%%%%%%%%%%%%%%%%%%%%%%%%%%%%%%%%%%%%%%%%%%%%%%%%%%%%%%%%%%%%%%%%%%%%%%%%%%%%%%%%%%%%%%%%%%%%%%%%%%%
\begin{figure}  %[!hb]
     \centering
      \vspace{-0.5cm}
     \includegraphics[scale=0.46]{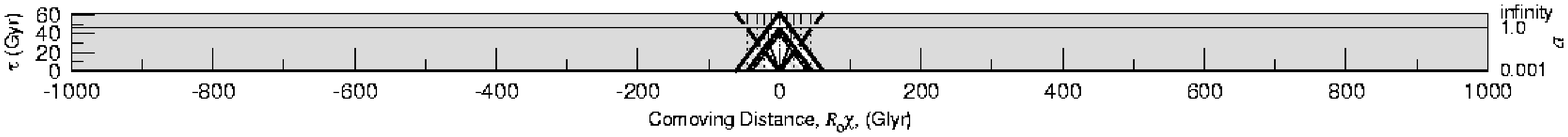}
      \vspace{-0.3cm}
     \caption{  }
     \label{fig:verywide}
      \vspace{-0.3cm}
\end{figure}
%%%%%%%%%%%%%%%%%%%%%%%%%%%%%%%%%%%%%%%%%%%%%%%%%%%%%%%%%%%%%%%%%%%%%%%%%%%%%%%%%%%%%%%%%%%%%%%%%%%%%%%%%%%%%%%%%%%%%%%
   
\subsection{Horizon problem}
\label{sec:horizonprob}

What should our assumptions be about regions of the Universe that have never been in causal contact?
If we look as far away as we can in one direction and as far away as we can in the other direction we can ask
the question, have those two points (points {\bf A} and {\bf B} in Fig.~\ref{fig:fig1bot}) been able to see each other.
In the standard big bang model without inflation the answer is no. Their past light cones are the little cones
beneath points {\bf A} and {\bf B}. Inserting a period of inflation during the early universe has the effect of moving
the surface of last scattering up to the line labeled ``new surface of last scattering''. Points {\bf A} 
and {\bf B} then
become points {\bf A'} and {\bf B'}. And the apexes of their past light cones are at points {\bf A'} and {\bf B'}. 
These two new light cones have a large degree of intersection. There would have been sufficient time
for thermal equilibrium to be established between these two points. Thus, the answer to the question: ``Why
are two points in opposite sides of the sky at the same temperature?'' is, because they have been in causal
contact and have reached thermal equilibrium. 

Five years ago most of us thought that as we waited patiently we would be rewarded with
a view of more and more of the Universe and eventually, we hoped to see the full extent of the inflationary bubble -- 
the size of the patch that inflated to form our Universe. However, $\Lambda$ has interrupted these dreams 
of unfettered empiricism.  
We now think there is an upper limit to the comoving size of the observable universe. 
In Fig. \ref{fig:fig1bot} we see that the observable universe ( = particle horizon) in the new 
standard $\Lambda$-CDM model approaches 62 billion light years in radius but will never extend further.
That is as large as it gets. That is as far as we will ever be able to see. Too bad.

\subsection{How big is a causally connected patch of the CMB without and with inflation?}
From Fig. \ref{fig:fig1bot} we can read off the x axis that the comoving radius of the base of the 
small light cone under points {\bf A} or {\bf B} is $r = R_{o}\chi \sim $ billion light years.
This is the current size of the patch that was causally connected at last scattering.
The physical size $D$ of the particle horizon today is $D \approx 47$ billion light years (Fig. \ref{fig:fig1bot}). 
The fraction $f$ of the sky occupied by one causally connected patch is $f = \pi r^{2}/ 4 \pi D^{2} \approx 1/9000$. 
The area of the full sky is about $40,000$ square
degrees ($ 4 \pi$ steradians). The area of a causally connected patch is 
$({\mbox \rm area \ of \ the\  sky})\times f  = 40,000/9,000 \approx 4$ square degrees.   % $\sim \pi 1^{2}$.

With inflation, the size of the causally connected patch depends on how many e-foldings of expansion occurred
during inflation. To solve the horizon problem we need a minimum of $\sim 60$ e-folds of expansion or an expansion 
by a factor of $\sim 10^{30}$. But since this is only a minimum, the full size of a causally connected patch, although
bigger than the observable universe, will never be known  unless it happens to be between $47$ Glyr (our current
particle horizon) and $62$ Glyr (the comoving size of our particle horizon at the end of time).

The constraint on the lower limit to the number of e-foldings $\sim 60$ 
(or $\sim 10^{30}$ ) comes from the requirement to solve the horizon problem.
What about the upper limit to the number of e-folds? How big is our inflationary bubble?
How big the inflationary patch is depends sensitively on 
when inflation happened, the height of the inflaton potential and how long 
inflation lasted ($t_{i},t_{e}$  and $\Lambda_{\rm inf}$ at Eq.~\ref{eq:last}) --
which in turn depends on the decay rate of the false vacuum. 
Without a proper GUT, these numbers cannot be approximated with any confidence. 
It is certainly reasonable to expect homogeneity to continue for some distance beyond our
observable universe but there does not seem to be any reason why it should go on forever.
In eternal inflation models, the homogeneity definitely does not go on forever (Liddle \& Lyth 2000).

When could inflation have occurred?
The earliest time is the Planck time at $10^{19}$ GeV or $10^{-43}$ seconds.
The latest is at the electroweak symmetry breaking at $10^{2}$ GeV or $10^{-12}$ seconds.  
The GUT scale is a favorite time at $10^{16}$ GeV or $10^{-35}$ seconds.
``Beyond these limits very little can be said for certain about inflation. So most papers
about inflationary models are more like historical novels than real history, and they
describe possible interactions that would be interesting instead of interactions that have
to occur. As a result, inflation is usually described as the inflationary scenario instead
of a theory or hypothesis. However, it seems quite likely that the inflation did occur, even
though we don't know when, or what the potential was.'' -- Wright (2003).     %p 72.

%%%%%%%%%%%%%%%%%%%%%%%%%%%%%%%%%%%%%%%%%%%%%%%%%%%%%%%%%%%%%%%%%%%%%%%%%%%%%%%%%%%%%%%%%%%%%%%%%%%%%%%%%%%%%%%%%%%%%%%
\begin{figure}[t!h]
%\figurebox{22pc}{15pc}{} % to have a box alone
\epsfxsize=32pc % will enlarge or reduce the postscript figures based on the xsize
\epsfbox{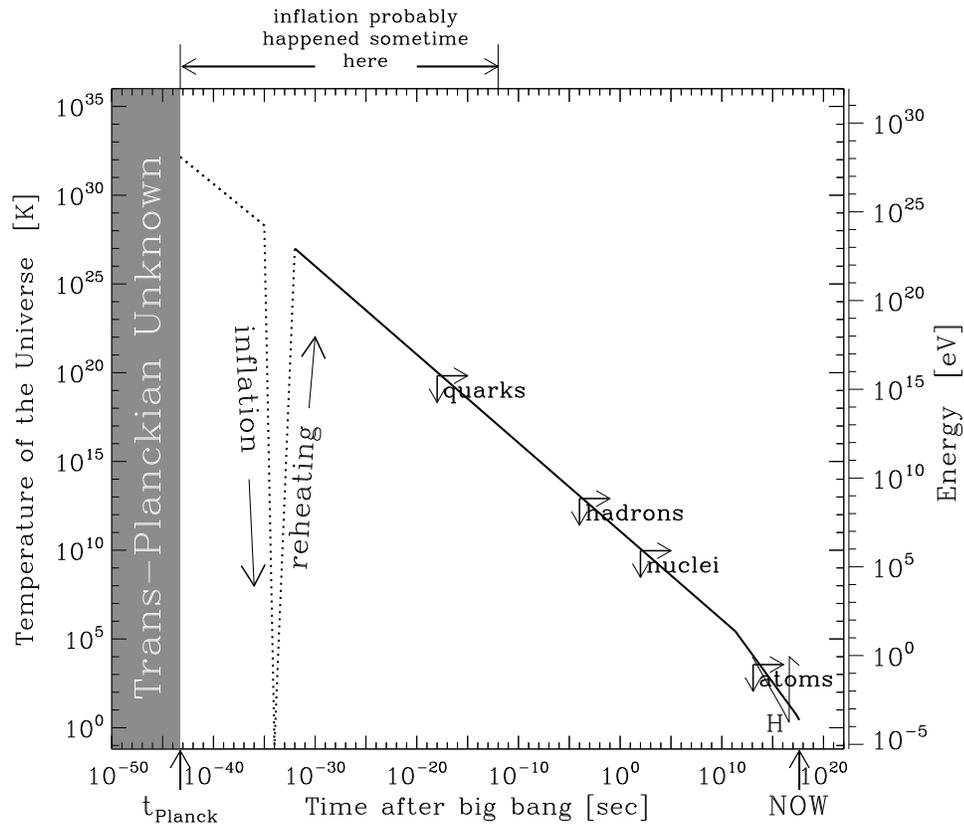} % postscript image file name   ; made by IDLpros/Schwartzman/Toft*.pro
\caption{{\bf Temperature of the Universe.}
The temperature and composition history of the standard big bang model with an epoch of
inflation and reheating inserted between $10^{-35}$ and $10^{-32}$ seconds after the
big bang. 
Inflation increases the size of the Universe, decreases the temperature and dilutes any structure.
Reheating then creates matter which decays and raises the temperature again.
This plot is also an overview of the energy scales at which the various components of our Universe froze out and
became permanent features. Quarks froze into protons and neutrons ($\sim$ GeV), protons and neutrons froze
into light nuclei ($\sim$ MeV), and these light nuclei froze into neutral atoms ($\sim$ eV) which cooled 
into molecules and then gravitationally collapsed into stars.
And now, huddled around these warm stars, we are living in the ice ages of the Universe with the CMB at 3K or $\sim 10^{-3}$ eV. 
}
\label{fig:Toft}
\end{figure}
%%%%%%%%%%%%%%%%%%%%%%%%%%%%%%%%%%%%%%%%%%%%%%%%%%%%%%%%%%%%%%%%%%%%%%%%%%%%%%%%%%%%%%%%%%%%%%%%%%%%%%%%%%%%%%%%%%%%%%%
%%%%%%%%%%%%%%%%%%%%%%%%%%%%%%%%%%%%%%%%%%%%%%%%%%%%%
\begin{figure}  %[!hb]
     \centering
    \includegraphics[angle=-90,scale=1.1]{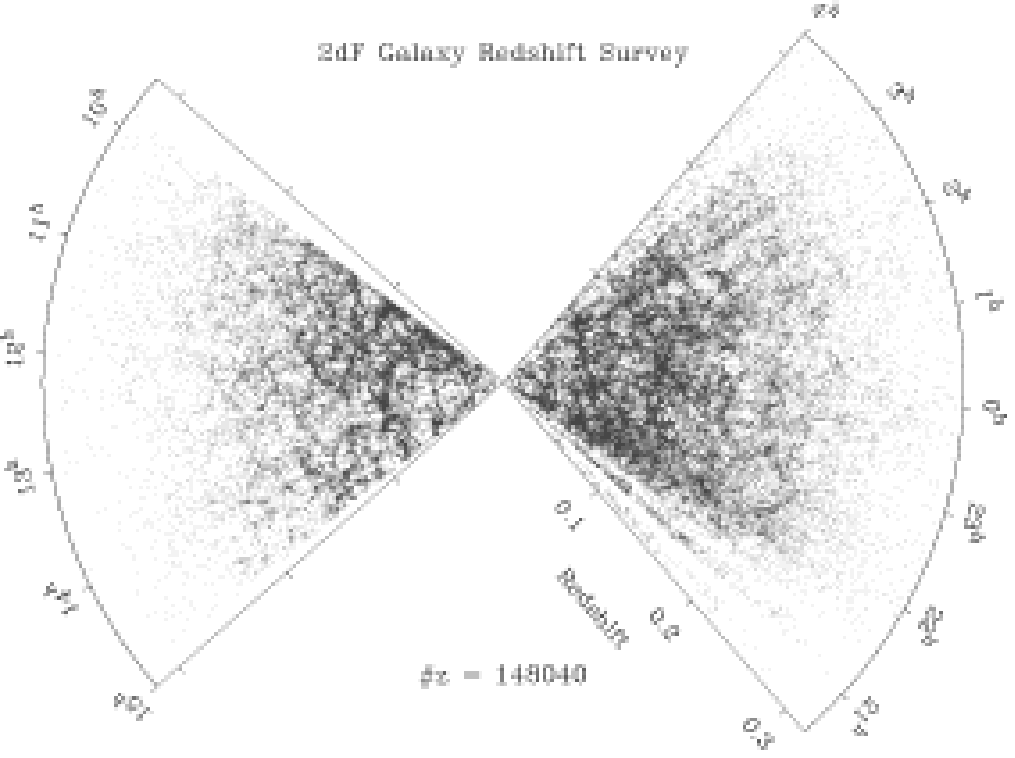}
    \includegraphics[angle=-90,scale=1.1]{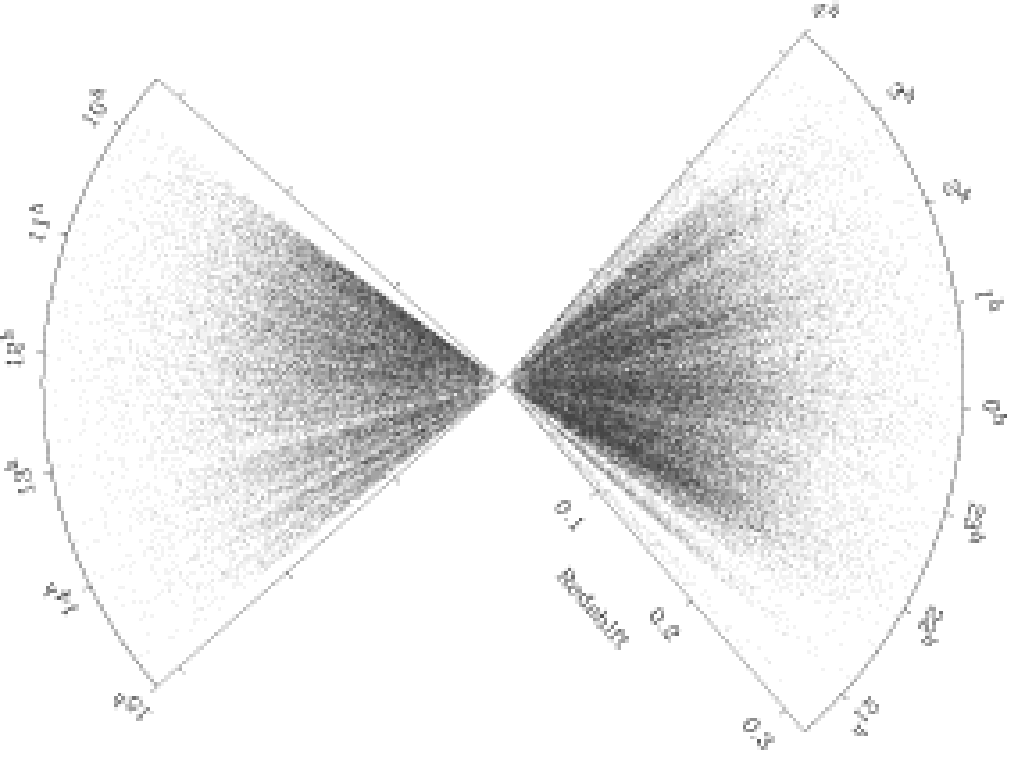}
    \caption{{\bf Real Structure (top) is not Random (bottom).} 
If galaxies were distributed randomly in the Universe with no 
large scale structure, the 2dF galaxy redshift survey of the Local Universe 
would have produced the lower map.
The upper map it did produce shows galaxies clumped into clusters 
radially smeared by the fingers of God, and 
empty voids surrounded by great walls of galaxies.
The same number of galaxies is shown in each panel.
Since all the large scale structure in the Universe has its origin in
inflation, we should be able to look at the details of this structure to
constrain inflationary models.
A minimalistic set of parameters to describe all this structure is the 
amplitude and the scale dependence of the density perturbations.
}
\label{fig:2df}
\end{figure}
%%%%%%%%%%%%%%%%%%%%%%%%%%%%%%%%%%%%%%%%%%%%%%%%%%%%

\section{How Does Inflation Produce All the Structure in the Universe?}
% One picosecond can make all the differences in the world.

In our Universe quantum fluctuations have been expanded into the largest structures we observe
and clouds of hydrogen have collapsed to form kangaroos. The larger end of this hierarchical range of structure --
the range controlled by gravity, not chemistry, is what inflation is supposed to explain.

Inflation produces structure because quantum mechanics, not classical mechanics describes the Universe
in which we live. 
The seeds of structure, quantum fluctuations, do not exist in a classical world.
If the world were classical, there would be no clumps or balls to populate classical mechanics textbooks.
Inflation dilutes everything -- all preexisting structure.  It empties the
Universe of anything that may have existed before, except quantum fluctuations.
These it can't dilute. These then become the seeds of who we are.

One of the most important questions in cosmology is:
What is the origin of all the galaxies, clusters, great walls, 
filaments and voids we see around us?
The inflationary scenario provides the most popular 
explanation for the origin of these structures: 
they used to be quantum fluctuations.
During the metamorphosis of quantum fluctuations
into CMB anisotropies and then into galaxies, primordial quantum fluctuations 
of a scalar field get amplified and evolve
to become classical seed perturbations and eventually
large scale structure.
{\it Primordial} quantum fluctuations are initial conditions. Like radioactive decay or quantum tunneling, 
they are not caused by any preceding event.

``Although introduced to resolve problems associated with the initial conditions needed for the Big Bang cosmology, inflation's
lasting prominence is owed to a property discovered soon after its introduction: It provides a possible explanation for the
initial inhomogeneities in the Universe that are believed to have led to all the structures we see, from the earliest objects
formed to the clustering of galaxies to the observed irregularities in the microwave background.''
-- Liddle \& Lyth (2000)  % p1 

%%%%%%%%%%%%%%%%%%%%%%%%%%%%%%%%%%%%
\begin{figure}[t]
%\figurebox{22pc}{15pc}{} 
    \centering
    \includegraphics[height=11cm]{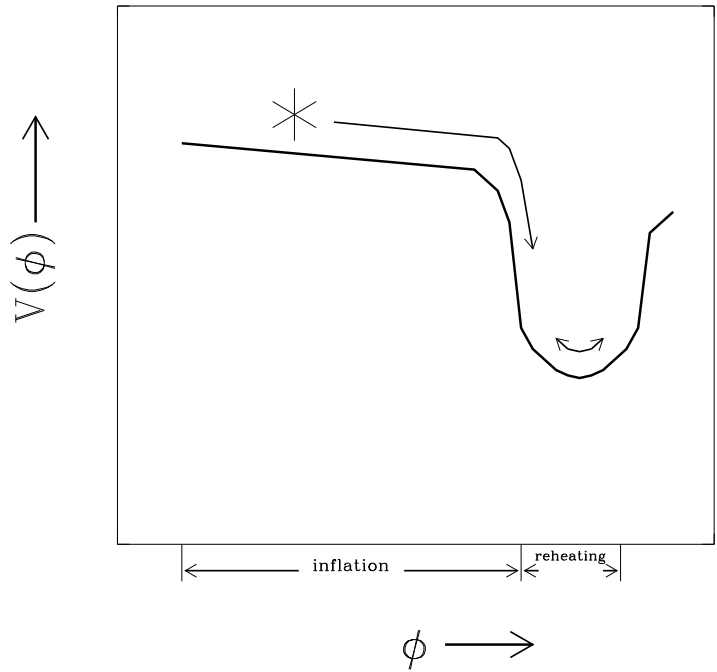}
    \caption{{\bf Model of the Inflaton Potential.}    %\\
%\epsfbox{v.ps}
A potential $V$ of a scalar field $\phi$ with a flat part and a valley.
The rate of expansion $H$ during inflation is related to the amplitude of the potential
during inflation. In the slow roll approximation $H^{2} = V(\phi)/m_{pl}^{2}$ (where $m_{pl}$ is the Planck mass).
Thus, from Eq.~\ref{eq:lamb} we have $\Lambda_{\rm inf} = 3\; V(\phi)/ m_{pl}^{2}$.
Thus, the height of the potential during inflation determines the rate of expansion during inflation.
And the rate at which the ball rolls  (the star rolls in this case) is determined by how steep the slope is: 
$\dot{\phi} = V^{\prime}/3H$. 
In modern physics, the vacuum is the state of lowest possible energy density.
The non-zero value of $V(\phi)$ is false vacuum -- a temporary state of lowest possible energy density.
The only difference between false vacuum and the cosmological constant is the stability of
the energy density -- how slow the roll is. Inflation lasts for $\sim 10^{-35}$ seconds while
the cosmological constant lasts $\gsim 10^{17}$ seconds. 
}
\label{fig:potent}
\end{figure}
%%%%%%%%%%%%%%%%%%%%%%%%%%%%%%%%%%%%

In early versions of inflation, it was hoped that the GUT scale Higgs potential could be used to
inflate. But the GUT theories had 1st order phase transitions. 
All the energy was dumped into the bubble walls and the observed structure in the Universe was supposed 
to come from bubble wall collisions. But the energy had to be spread out evenly. Percolation
was a problem and so too was a graceful exit from inflation.
New Inflation involves second order phase transitions (slow roll approximations).
The whole universe is one bubble and structure cannot come from collisions.
It comes from quantum fluctuations of the fields.
There is one bubble rather than billions and the energy gets dumped everywhere, not just at the bubble wall.

One way to understand how quantum fluctuations become real fluctuations is this.
Quantum fluctuations, i.e., virtual particle pairs of borrowed energy $\Delta E$, 
get separated during the interval $\Delta t \lsim \hbar/\Delta E$.
The $\Delta x$ in $\Delta x \lsim \hbar/\Delta p$ is a measure of their separation.
If during $\Delta t$ the physical size $\Delta x$ leaves the event horizon, the virtual particles cannot
reconnect, they become real and the energy debt must be paid by the driver of inflation, the energy of
the false vacuum -- the $\Lambda_{\rm inf}$ associated with the inflaton potential $V(\phi)$
(see Fig.~\ref{fig:potent}).

What kind of choices does the false vacuum have when it decays?
If there are many pocket universes, what are they like? Do they have the same value for the speed of
light? Are their true vacua the same as ours? Do the Higgs fields give the particles and forces the 
same values that reign in our Universe? Is the baryon asymmetry the same as in our Universe?

%%%%%%%%%%%%
\section{The Status of Inflation}
\label{sec:status}

Down to Earth astronomers are not convinced that inflation is a useful model. 
For them, inflation is a cute idea that takes a geometric flatness problem and replaces it with an inflaton potential
flatness problem.  It moves the problem to earlier times, it does not solve it. 
Inflation doesn't solve the fine-tuning problem. It moves the problem from ``Why is the Universe so
flat?'' to ``Why is the inflaton potential so flat?''. When asked, 
``Why is the Universe so flat?'', Mr Inflation responds, ``Because my inflaton potential is so flat.''
``But why is your inflaton potential so flat?''
``I don't know. It's just an initial condition.'' 
This may or may not be progress.
If we are content to believe that spatial flatness is less fundamental than inflaton 
potential flatness then we have made progress.

\subsection{Inflationary Observables}

Models of inflation usually consist of choosing a form for the potential $V(\phi)$.
A simple model of the potential is $V(\phi) = m^{2} \phi^{2}/2$ where the derivative with
respect to $\phi$ is $V^{\prime} =  m^{2} \phi$
and  $V^{\prime\prime} =  m^{2}$. This leads to a prediction for the observable spectral 
index of the CMB power spectrum:  $n_{s} = 1- 8m_{p}/\phi^{2}$ (e.g. Liddle \& Lyth 2000).
Estimates of the slope of the CMB power spectrum $n_{s}$ and its derivative $\frac{dn_{s}}{dk}$
have begun to constrain models of the inflaton potential (Table 1 and Spergel~2003).

The observational scorecard of inflation is mixed. 
Based on inflation, many theorists became convinced 
that the Universe was spatially flat despite many measurements to the contrary. 
The Universe has now been measured to be flat to high precision -- score one for inflation.
Based on vanilla inflation, most theorists thought that the flatness 
would be without $\Lambda$ -- score one for the observers.
Guth wanted to use the Georgi-Glashow GUT model as the potential to form structure. It didn't work -- score
one against inflation. But other plausible inflaton potentials can work.
Inflation seems to be the only show in town as far as producing
the seeds of structure -- score one for inflation.
Inflation predicts the spectral index of CMB fluctuations to be 
$n_{s} \approx 1$ -- score one for inflation.
But we knew that  $n_{s} \approx 1$ before inflation (minus 1/2 point for cheating). 
So far most of inflation's predictions have been retrodictions -- explaining
things that it was designed to explain.

Inflationary models and the new ekpyrotic models  %\cite{Ekpyrot} 
make different predictions about the slope $n_{T}$ %and amplitude 
of the tensor mode contribution to the CMB power spectrum.
Inflation has higher amplitudes at large angular scales  while ekpyrotic models
have the opposite. However, since the amplitude $T$ is unknown, finding
the ratio of the amplitude of tensor to scalar modes, $r=T/S \sim 0$, 
does not really distinguish the two models. 
Finding a value $ r > 0$ would however be interpreted as favoring 
inflation over ekpyrosis. 
Recent WMAP measurements of the CMB power
spectrum yield  $r <0.71$ at the 95\%  confidence level.

Measurements of CMB polarization over the next five years
will add more diagnostic power to CMB parameter estimation
and may be able to usefully constrain the slope and amplitude of tensor modes if they exist
at a detectable level.

One can be skeptical about the status of the problems that inflation claims to have solved.
After all, the electron mass is the same everywhere. 
The constants of nature are the same everywhere. 
The laws of physics seem to be the same everywhere.
If these uniformities need no explanation then why should the uniform temperatures, flat geometry and seeds of 
structure need an explanation.
Is this first group more fundamental than the second?

The general principle seems to be that 
if we can't imagine plausible alternatives then 
no explanation seems necessary. Thus, dreaming up imaginary alternatives creates imaginary problems, to which imaginary
solutions can be devised, whose explanatory power depends on 
whether the Universe could have been other than what it is.
However, it is not easy to judge the reality of counterfactuals.
Yes, inflation can cure the initial condition ills of the standard big bang model, but is inflation
a panacea or a placebo?

Inflation is not a theory of everything.
It is not based on M-theory or any candidate for
a theory of everything. It is based on a scalar field.
The inflation may not be due to a scalar field $\phi$ and its potential $V(\phi)$.
Maybe it has more to do with extra-dimensions?

%%%%%%%%%%%%%%%%%%%%%%%%%%%%%%%%%%%%%%%%%%%%%%%%%%%%%%%%%%%%%%%%%%%%%%%%%%%%%%%%%%%%%%%%%%%%%%%%%%%%%%%%%%%%%%%
\clearpage
\section{CMB}

\subsection{History}
By 1930, the redshift measurements of Hubble and others had convinced many scientists that the Universe
was expanding. This  suggested that in the distant past the Universe was smaller and hotter.
In the 1940's an ingenious nuclear physicist George Gamow,  began to take the 
idea of a very hot early universe seriously, and with Alpher and Herman, began 
using the hot big bang model to try to 
explain the relative abundances of all the elements. 
Newly available nuclear cross-sections made the
calculations precise. Newly available computers made the calculations doable. In 1948 Alpher and Herman
published an article predicting that the temperature of the bath of photons left from the early universe
would be 5 K. They were told by colleagues that the detection of such a cold ubiquitous signal would be impossible.

In the early 1960's, Arno Penzias and Robert Wilson discovered excess antenna noise in a
horn antenna at Crawford Hill, Holmdel, New Jersey. They didn't know what to make of it.
Maybe the white dielectric material left by pigeons had something to do with it? 
During a plane ride, Penzias explained his excess noise problem to a fellow radio astronomer
Bernie Burke. Later, Burke heard about a talk by a young Princeton post-doc named Peebles, describing
how Robert Dicke's Princeton group was gearing up to measure radiation left over from an earlier hotter phase of
the Universe. Peebles had even computed the temperature to be about 10 K (Peebles 1965). 
Burke told the Princeton group about Penzias and Wilson's noise and Dicke gave Penzias a call.

Dicke did not like the idea that all the matter in the Universe had been created in the big bang. 
He liked the oscillating universe. He knew however that the first stars had fewer heavy elements. Where were
the heavy elements that had been produced by earlier oscillations? -- these elements must have been destroyed
by the heat of the last contraction. Thus there must be a remnant of that heat and Dicke had decided to look for it.
Dicke had a theory but no observation to support it. Penzias had noise but no theory. After the phone call
Penzias' noise had become Dicke's observational support.

Until 1965 there were two competing paradigms to describe the early universe: the big bang model and the steady
state model. 
The discovery of the CMB removed the steady state model as a serious contender. The big bang model had predicted
the CMB; the steady state model had not.

\subsection{What is the CMB?}

The observable universe is expanding and cooling. Therefore in the past it was hotter and smaller. The cosmic microwave 
background (CMB) is the after glow of thermal radiation left over from this hot early epoch in the evolution of
the Universe. It is the redshifted relic of the  hot big bang.
 The CMB is a bath of photons coming from every direction. These are the oldest photons one can observe
 and they contain information about the Universe at redshifts much larger than the redshifts of galaxies and quasars
($z \approx 1000 >> z \approx$ few).

Their long journey toward us has lasted more than 99.99\% of the 
age of the Universe and began when the Universe
was one thousand times smaller than it is today.
The CMB was emitted by the hot plasma of the Universe long before 
there were planets, stars or galaxies.
The CMB is thus a unique tool for probing the early universe.

One of the most recent and most important advances in astronomy has been the discovery 
of hot and cold spots in the CMB based on data from the COBE satellite (Smoot et al. 1992).
This discovery has been hailed as ``Proof of the Big Bang''
and the ``Holy Grail of Cosmology'' and elicited comments like:
``If you're religious it's like looking at the face of God''
(George Smoot) and
``It's the greatest discovery of the century, if not of all time''
(Stephen Hawking).
As a graduate student analyzing COBE data at the time,
I knew we had discovered something fundamental but
its full import didn't sink in until one night after
a telephone interview for BBC radio. I asked the interviewer for
a copy of the interview, and he told me that would be
possible if I sent a request to the {\it religious affairs} department.

\begin{figure}[t]
%\figurebox{22pc}{15pc}{} % to have a box alone
%\centerline{\psfig{figure=screaming6.eps,angle=0,height=10cm,width=11cm,rheight=11cm,rwidth=11cm}}
\epsfxsize=29pc % will enlarge or reduce the postscript figures based on the xsize
\epsfbox{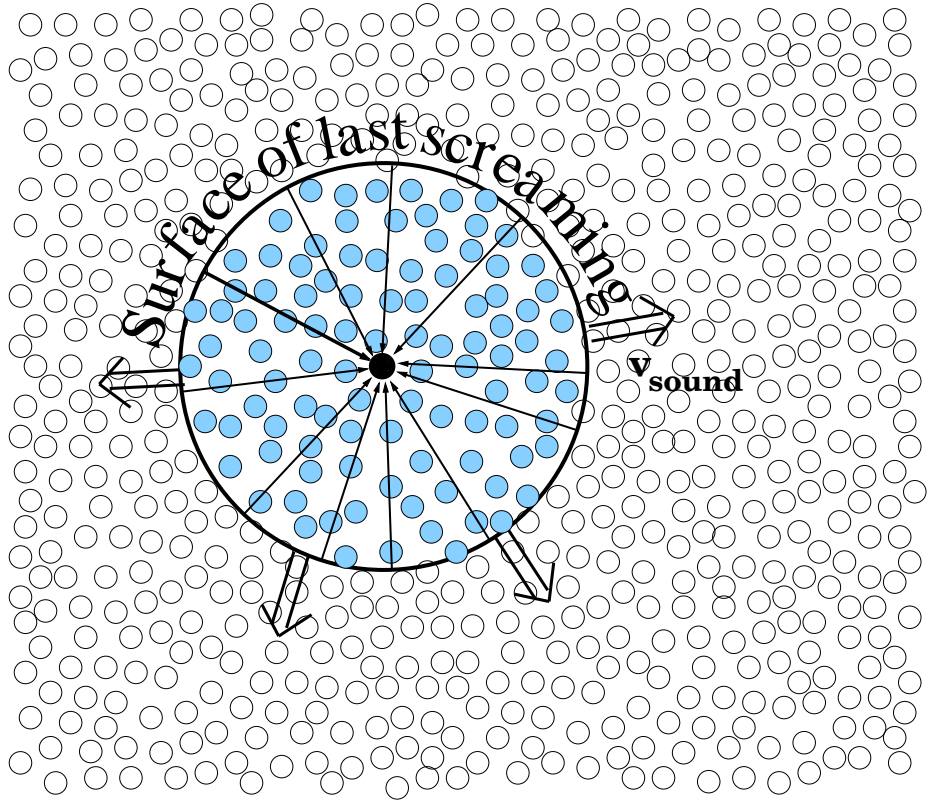} % postscript image file name
\caption{{\bf The Surface of Last Screaming.}
Consider an infinite field full of people screaming. 
The circles are their heads. 
You are screaming too. (Your head is the black dot.)
Now suppose everyone stops screaming at the same time. 
What will you hear? Sound travels at 330 m/s.
One second after everyone stops screaming you will be able to hear the screams from a 
`surface of last screaming' 330 meters away from you in all directions.
After 3 seconds the faint screaming will
be coming from 1 km  away...etc. 
No matter how long you wait, faint screaming will always be coming from 
the surface of last screaming -- a surface that is receding from you
at the speed of sound (`v$_{sound}$').
The same can be said of any observer -- each is the center of a surface
of last screaming.
In particular, observers on your surface of last screaming are currently hearing you scream
since you are on their surface of last screaming.
The screams from the people closer to you than the surface of last
screaming  have passed you by -- you hear nothing from them (gray heads).
When we observe the CMB in every direction
we are seeing photons from the surface of last scattering. 
We are seeing back to a time soon after the big bang when the entire
universe was opaque (screaming).
}
\label{fig:lastscreaming}
\end{figure}

The CMB comes from the surface of last scattering of the Universe.
When you look into a fog, you are looking at a surface of
last scattering. It is a surface defined by all the molecules
of water which scattered a photon into your eye.
On a foggy day you can see 100 meters, on really foggy days
you can see 10 meters. If the fog is so dense you cannot see 
your hand then the surface of last scattering 
is less than an arm's length away.
Similarly, when you look at the surface of the Sun you are seeing
photons last scattered by the hot plasma of the photosphere.
The early universe is as hot as the Sun and similarly
the early universe has a photosphere (the surface of last scattering) 
beyond which (in time and space) we cannot see. 
As its name implies, the surface of last scattering is where
the CMB photons were scattered for the last time before
arriving in our detectors.
The `surface of last screaming' presented in Fig. \ref{fig:lastscreaming} 
is a pedagogical analog.

%%%%%%%%%%%%%%%%%%%%%%%%%%%%%%%%%%%%
%%%%%%%%%%%%%%%%%%%%%%%%%%%%%%%%%%%%

\subsection{Spectrum}

The big bang model predicts that the cosmic background radiation will be thermalized -- it will have a blackbody
spectrum. The measurements of the antenna temperature of the radiation at various frequencies between 1965 and
1990 had shown that the spectrum was approximately blackbody but there were some measurements at high frequencies
that seemed to indicate an infrared excess --  a bump in the spectrum that was not easily explained.  
In 1989, NASA launched the COBE (Cosmic Background Explorer) satellite to investigate the cosmic microwave and infrared
background radiation. There were three instruments on board. After one year of observations the FIRAS instrument 
had measured the spectrum of the CMB  and found it to be a blackbody spectrum. The most recent analysis
of the FIRAS data gives a temperature of $2.725 \pm 0.002$ K (Mather et al. 1999). 

A CMB of cosmic origin
(rather than one generated by starlight processed by iron needles in the intergalactic medium) 
is expected to have a blackbody spectrum 
and to be extremely isotropic.
COBE FIRAS observations show that the CMB is very well approximated by an isotropic blackbody.

\subsection{Where did the energy of the CMB come from?}

Recombination occurs when the CMB temperature has dropped low enough such that
there are no longer enough high energy
photons to keep hydrogen ionized; $\gamma + H \leftrightarrow e^{-} + p^{+}$.
Although the ionization potential of hydrogen is 13.6 eV ($T \sim 10^{5}$ K), recombination 
occurs at $T \approx 3000$ K.  This low temperature can be explained by the fact that 
there are a billion photons for every proton in the Universe. This allows 
the high energy tail of the Planck distribution of the photons to keep
the comparatively small number of hydrogen atoms ionized until temperatures and energies much lower than 13.6 eV.
The Saha equation (e.g. Lang 1980) describes this balance between the ionizing photons and the ionized and neutral hydrogen.

The energy in the CMB did not come from the recombination of electrons with protons to form hydrogen at the surface
of last scattering. That contribution is negligible -- only about one 10 eV photon for each baryon, while there are
$\sim 10^{10}$ times more CMB photons than baryons and each of those photons {\it at recombination} had an energy of 
$\sim 0.3$ eV: $\frac{\Delta E_{rec}}{E_{CMB}} = \frac{10 eV \times 10^{-10}}{0.3 eV} \sim 10^{-9}$. 
The energy in the CMB came from the annihilation of particle/anti-particle pairs during a very early epoch 
called baryogenesis and 
later when electrons and positrons annihilated at an energy of $\sim 1$ MeV.

As an example of energy injection, consider the thermal bath of neutrinos that fills the Universe. It  decoupled 
from the rest of 
the Universe at an energy above an MeV. After decoupling the neutrinos and the photons, both being relativistic, 
cooled as $T \propto R^{-1}$.
If nothing had injected energy into the Universe below an MeV, the neutrinos and the photons would both have
a temperature today of $1.95$ K. However the photons have a temperature of $2.725$ K. Where did this 
extra energy come from? It came from the annihilation of electrons and positrons when the temperature of the Universe
fell below an MeV. This process injected energy into the Universe by heating up the residual electrons, which in turn heated up
the CMB photons. The relationship between the CMB and neutrino temperatures is $T_{CMB} = (11/4)^{1/3}\; T_{\nu}$.
Derivation of this result using  entropy conservation during electron/positron annihilation can be found in
Wright (2003) %p 27, 
or Peacock (2000).
The bottom line: $T_{CMB}= 2.7$ K $>  T_{\nu} = 1.9$ K  because the photons were heated up by $e^{\pm}$ annihilation 
while the neutrinos were not.
This temperature for the neutrino background has not yet been confirmed observationally.

\subsection{Dipole}

To a very good approximation the CMB is a flat featureless blackbody; there are no anisotropies and the temperature
is a constant $T_{o} = 2.725$ K in every direction. When we remove this mean value, the next largest feature visible at 
1000 times smaller amplitude is the kinetic dipole.
Just as the 17 satellites of the Global Positioning System (GPS) provide a reference frame to establish positions and 
velocities on the Earth.
The CMB gives all the inhabitants of the Universe a special common rest frame with respect to which 
all velocities can be measured -- the comoving frame in which the observers  see no CMB dipole.  
People who enjoy special relativity but not general relativity often baulk at this concept.
A profound question that may make sense is: Where did the rest frame of the CMB come from? How was it chosen?
Was there a mechanism for a choice of frame, analogous to the choice of vacuum during spontaneous symmetry breaking?

%%%%%%%%%%%%%%%%%%%%%%%%%%%%%%%%%%%%
\begin{figure}[t]
%\figurebox{22pc}{15pc}{} % to have a box alone
\epsfxsize=34pc % will enlarge or reduce the postscript figures based on the xsize
\epsfbox{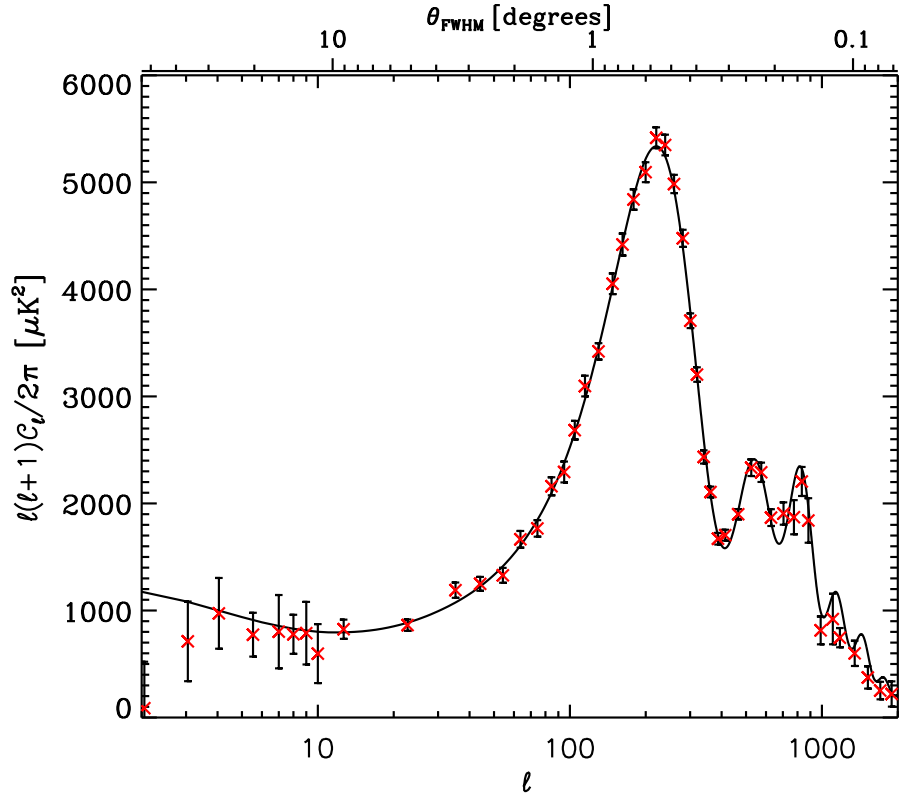} % postscript image file name
\caption{{\bf Measurements of the CMB power spectrum.}
CMB power spectrum from the world's combined data, including the recent WMAP satellite results
(Hinshaw et al. 2003).
The amplitudes of the hot and cold spots in the CMB depend on their angular
size. Angular size is noted in degrees on the top x axis.
% and with the Legendre index $\ell$ on the bottom x axis. 
The y axis is the power in the temperature fluctuations.
No CMB experiment is sensitive to this entire range of angular scale.
When the measurements at various angular scales are put together they form the CMB power spectrum.
At large angular scales ($\ell \lsim 100$), the temperature fluctuations are on 
scales so large that they are `non-causal', i.e., they have physical sizes 
larger than the distance light could have traveled between the big bang (without inflation) and their 
age at the time we see them (300,000 years after the big bang).
They are either the initial conditions of the Universe or were laid down
during an epoch of inflation $\sim 10^{-35}$ seconds after the big bang. 
New data are being added to these points every few months. 
The concordance model shown has the following cosmological parameters:  $\oll = 0.743$,
$\Omega_{CDM} = 0.213$, $\Omega_{baryon} = 0.0436$, $h=0.72$, $n=0.96$, $\tau = 0.12$ and
no hot dark matter (neutrinos) ($\tau$ is the optical depth to the surface of last scattering).
$\chi^{2}$ fits of this data to such model curves yields the estimates in Table 1.
The physics of the acoustic peaks is briefly described in Fig. \ref{fig:acoustic}.
}
\label{fig:binnedcls}
\end{figure}
%%%%%%%%%%%%%%%%%%%%%%%%%%%%%%%%%%%%

\subsection{Anisotropies}

Since the COBE discovery of hot and cold spots in the CMB, 
anisotropy detections have been reported 
by more than two dozen groups with various instruments, at various frequencies and 
in various patches and swathes of the microwave sky. 
Figure \ref{fig:binnedcls} is a compilation 
of the world's measurements (including the recent WMAP results).
Measurements on the left (low $\ell$'s) are at large 
angular scales while most recent measurements are trying to constrain 
power at small angular scales.
The dominant peak at $\ell \sim 200$ and the smaller amplitude peaks at smaller angular scales are due
to acoustic oscillations in the photon-baryon fluid in cold dark matter gravitational potential wells and hills.
%(Hu \etal 1996, Lineweaver 1999b).
The detailed features of these peaks in the power spectrum are dependent on a large number  of
cosmological parameters.

%%%%%%%%%%%%%%%%%%%%%%%%%%%%%%%%%%%%
\begin{figure}  %[t]
%\figurebox{22pc}{15pc}{} % to have a box alone
\epsfxsize=25pc % will enlarge or reduce the postscript figures based on the xsize
\epsfbox{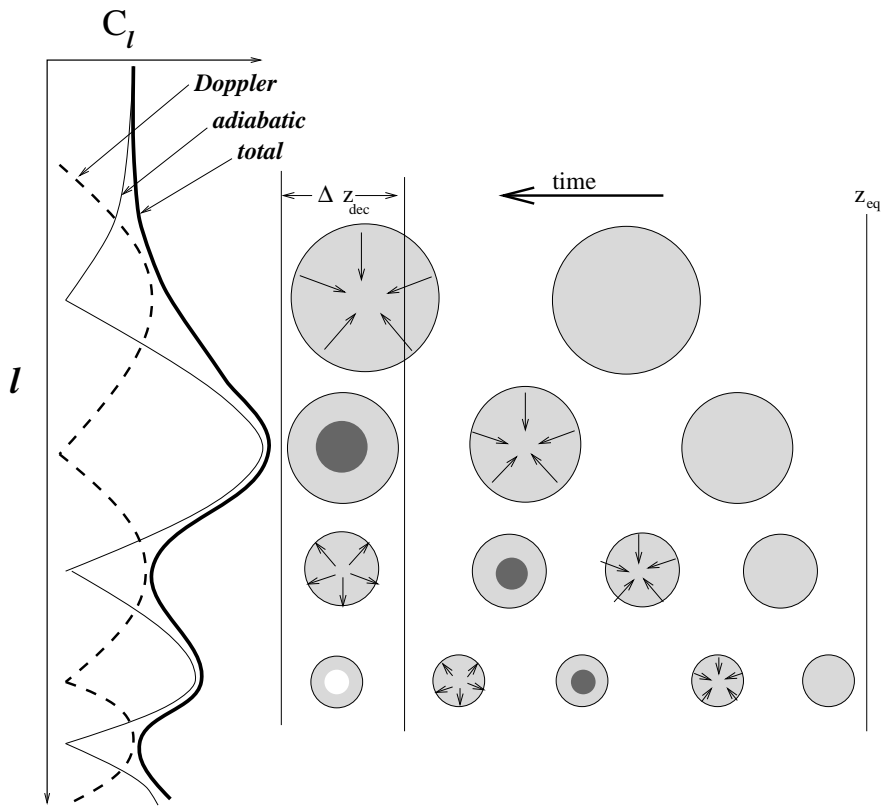} % postscript image file name
\caption{The dominant acoustic peaks in the CMB power spectra are caused by the collapse of 
dark matter over-densities and the
oscillation of the photon-baryon fluid into and out of these over-densities.
After matter becomes the dominant component of the Universe, at $z_{eq} \approx 3233$ (see Table 1), 
cold dark matter potential wells (gray spots) 
initiate in-fall and then oscillation of the photon-baryon fluid. 
The phase of this in-fall and oscillation at $z_{dec}$ (when photon pressure disappears) determines
the amplitude of the power as a function of angular scale. The bulk motion of the photon-baryon
fluid produces `Doppler' power out of phase with the adiabatic power. The power spectrum (or $C_{\ell}$s)
is shown here rotated by $90 \dg$ compared to Fig. \ref{fig:binnedcls}.
Oscillations in fluids are also known as sound.
Adiabatic compressions and rarefactions become visible in the radiation when the baryons 
decouple from the photons during the interval marked $\Delta z_{dec}$ ($\approx 195 \pm 2$, Table 1).
The resulting bumps in the power spectrum are analogous to the standing waves of 
a plucked string. This very old music, when converted
into the audible range, produces an interesting roar (Whittle 2003).
Although the effect of over-densities is shown, we are in the linear regime so under-densities contribute
an equal amount. That is, each acoustic peak in the power spectrum is made of equal contributions
from hot and cold spots in the CMB maps (Fig. \ref{fig:BWcleaned512}).
Anisotropies on scales smaller than about $8'$ are suppressed because they are superimposed on each other over
the finite path length of the photon through the surface $\Delta z_{dec}$.
%See Hu {\it etal} (1997) and Lineweaver (1997) for details.
}
\label{fig:acoustic}
\end{figure}
%%%%%%%%%%%%%%%%%%%%%%%%%%%%%%%%%%%%
%%%%%%%%%%%%%%%%%%%%%%%%%%%%%%%%%%%%
\begin{figure}  %[b!h]
%\figurebox{22pc}{15pc}{} % to have a box alone
%\epsfxsize=3.0pc % will enlarge or reduce the postscript figures based on the xsize
%\epsfbox{TegBW512.ps} % postscript image file name
\includegraphics[angle=90,scale=1.0]{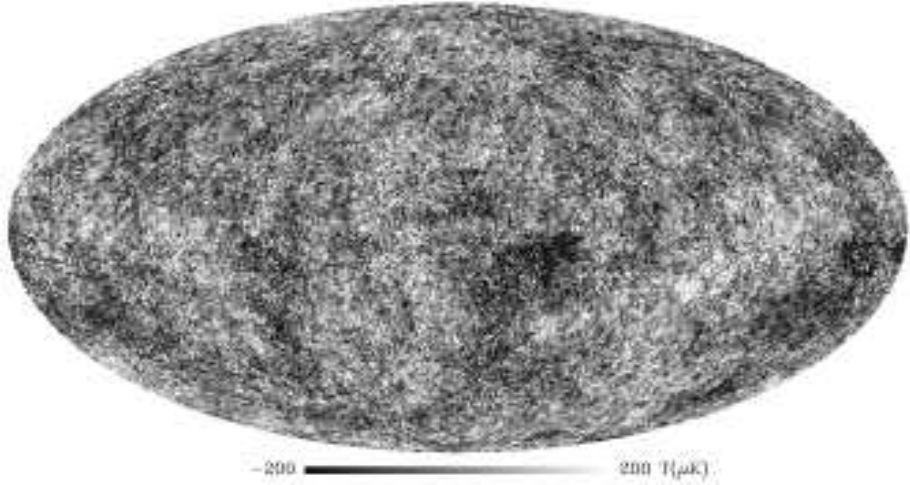}
\caption{Full sky temperature map of the cosmic microwave background derived from the
WMAP satellite (Bennett et al 2003, Tegmark et al 2003). The disk of the Milky Way runs
horizontally through the center of the image but has been almost completely removed from 
this image.
The angular resolution of this map is about 20 times better than its predecessor, the COBE-DMR
map in which the hot and cool spots shown here were detected for the first time.
The large and small scale power of this map is shown separately in the next figure.
}
\label{fig:BWcleaned512}
\end{figure}
%%%%%%%%%%%%%%%%%%%%%%%%%%%%%%%%%%%%
%\clearpage
%%%%%%%%%%%%%%%%%%%%%%%%%%%%%%%%%%%%
\begin{figure}  %[t!h]
%\figurebox{22pc}{15pc}{} % to have a box alone
%\epsfxsize=3.0pc % will enlarge or reduce the postscript figures based on the xsize
%\epsfbox{TegBW512.ps} % postscript image file name
\includegraphics[angle=90,scale=1.0]{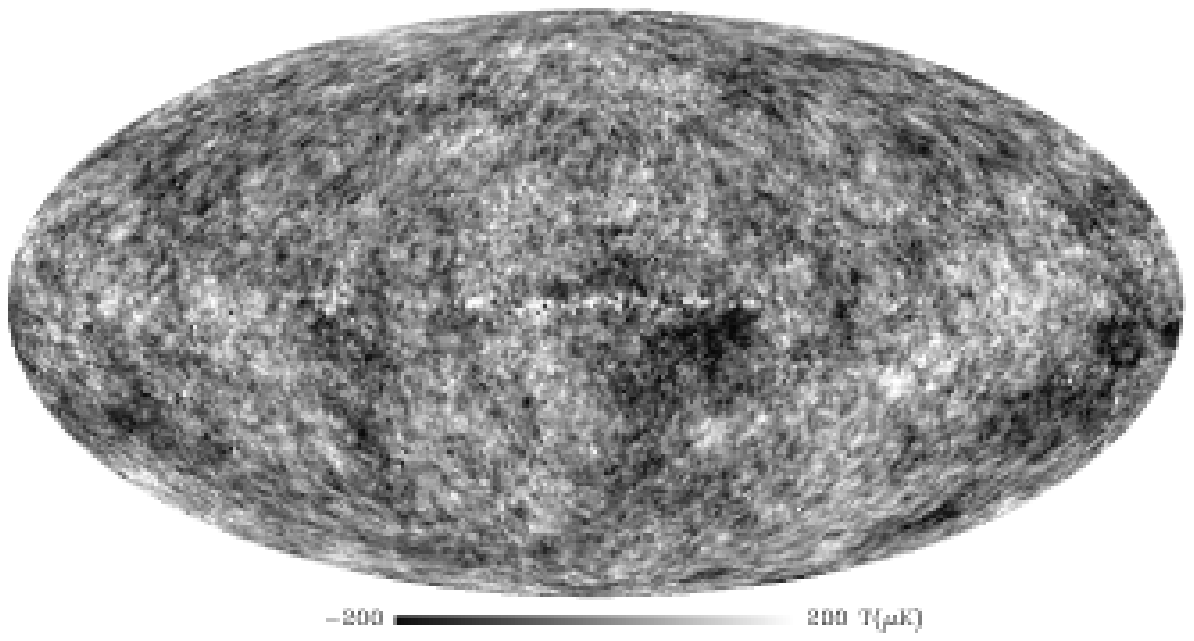}
\includegraphics[angle=90,scale=1.0]{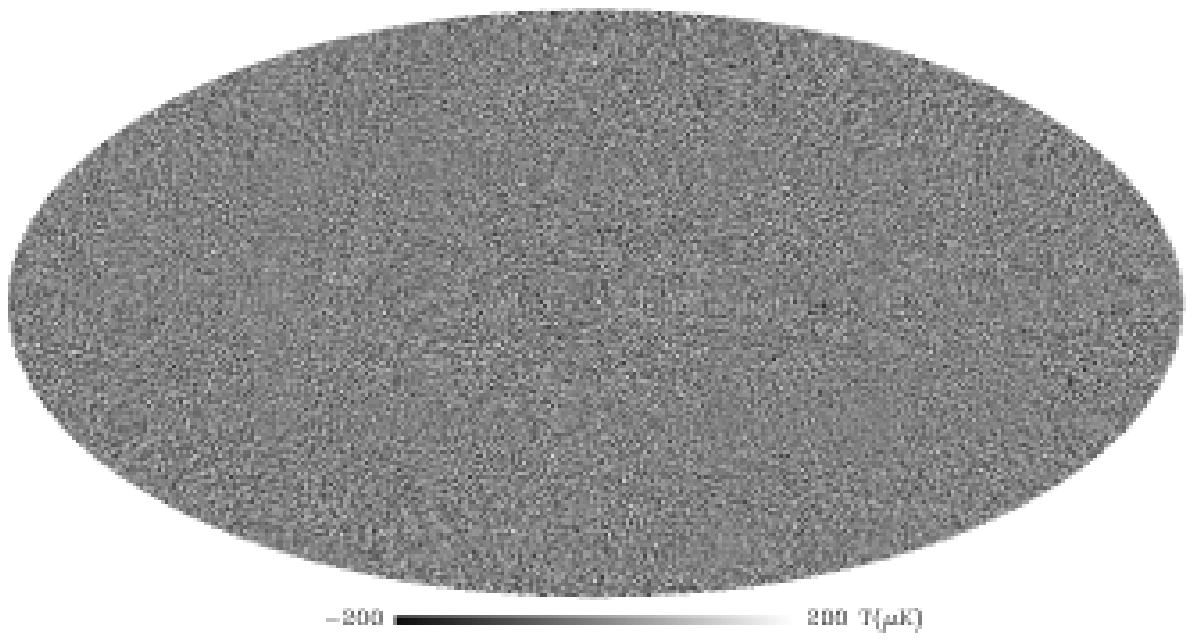}
\caption{Two basic ingredients: old quantum fluctuations (top) and new sound (bottom). 
These two maps were constructed from Fig. \ref{fig:BWcleaned512}. The top
map is a smoothed version of Fig. \ref{fig:BWcleaned512} and shows only power
at angular scales greater than $\sim 1 \deg$ ($\ell \lsim 100$, see Fig. \ref{fig:binnedcls}).
This footprint of the inflationary epoch was made in the first picosecond after the big
bang. In the standard big bang without inflation, all the structure here has to be attributed
to initial conditions.
The lower map was made by subtracting the top map from Fig. \ref{fig:BWcleaned512}.
That is, all the large scale power was subtracted from the CMB leaving only the 
small scale power in the acoustic peaks ($\ell > 100$, see Fig.~\ref{fig:binnedcls}) --
these are the crests of the sound waves generated after radiation/matter
equality (Fig.~\ref{fig:acoustic}).
Thus, the top map shows quantum fluctuations imprinted when the age of the Universe 
was in the range $[10^{-43}, 10^{-12}]$ seconds old, 
while the bottom map shows foreground contamination from sound generated when 
the Universe was $\sim 10^{13}$ seconds old. 
}
%(Healpix resolution 0.9 degrees, Nside = 512)
% high resolution is 7 arcminutes
\label{fig:twomaps}
\end{figure}
%%%%%%%%%%%%%%%%%%%%%%%%%%%%%%%%%%%%

\subsection{What are the oldest fossils we have from the early universe?}
It is sometimes said that the CMB gives us a glimpse of the Universe when it was $\sim 300,000$ years old.
This is true but it also gives us a glimpse of the Universe when it was less than a trillionth of
a second old. The acoustic peaks in the power spectrum (the spots of size less than about 1 degree) come from
sound waves in the photon-baryon plasma at $\sim 300,000$ years after the big bang but there is much structure
in the CMB on angular scales greater than 1 degree. When we look at this structure we are looking at the Universe
when it was less than a trillionth of a second old. 
The large scale structure on angular scales greater than $\sim 1$ degree is the oldest fossil
we have and dates back to the time of inflation. In the standard big bang model, structure on these acausal scales
can only be explained with initial conditions.

The large scale features in the CMB, i.e., all the features in the top
map of Fig.~\ref{fig:twomaps} but none of the features in the lower map,
are the largest and most distant objects every seen.
And yet they are probably also the smallest for they are quantum fluctuations
zoomed in on by the microscope called inflation and hung up in the sky.
So this map belongs in two different sections of the Guinness book 
of world records.

The small scale structure  on angular scales less than $\sim 1$ degree (lower map) results from
oscillations in the photon-baryon fluid between the redshift of equality and recombination.
Figure \ref{fig:acoustic} describes these oscillations in more detail.

%%%%%%%%%%%%%%%%%%%%%%%%%%%%%%%%%%%%
\begin{figure}[t]
%\figurebox{22pc}{15pc}{} % to have a box alone
\epsfxsize=32pc % will enlarge or reduce the postscript figures based on the xsize
\epsfbox{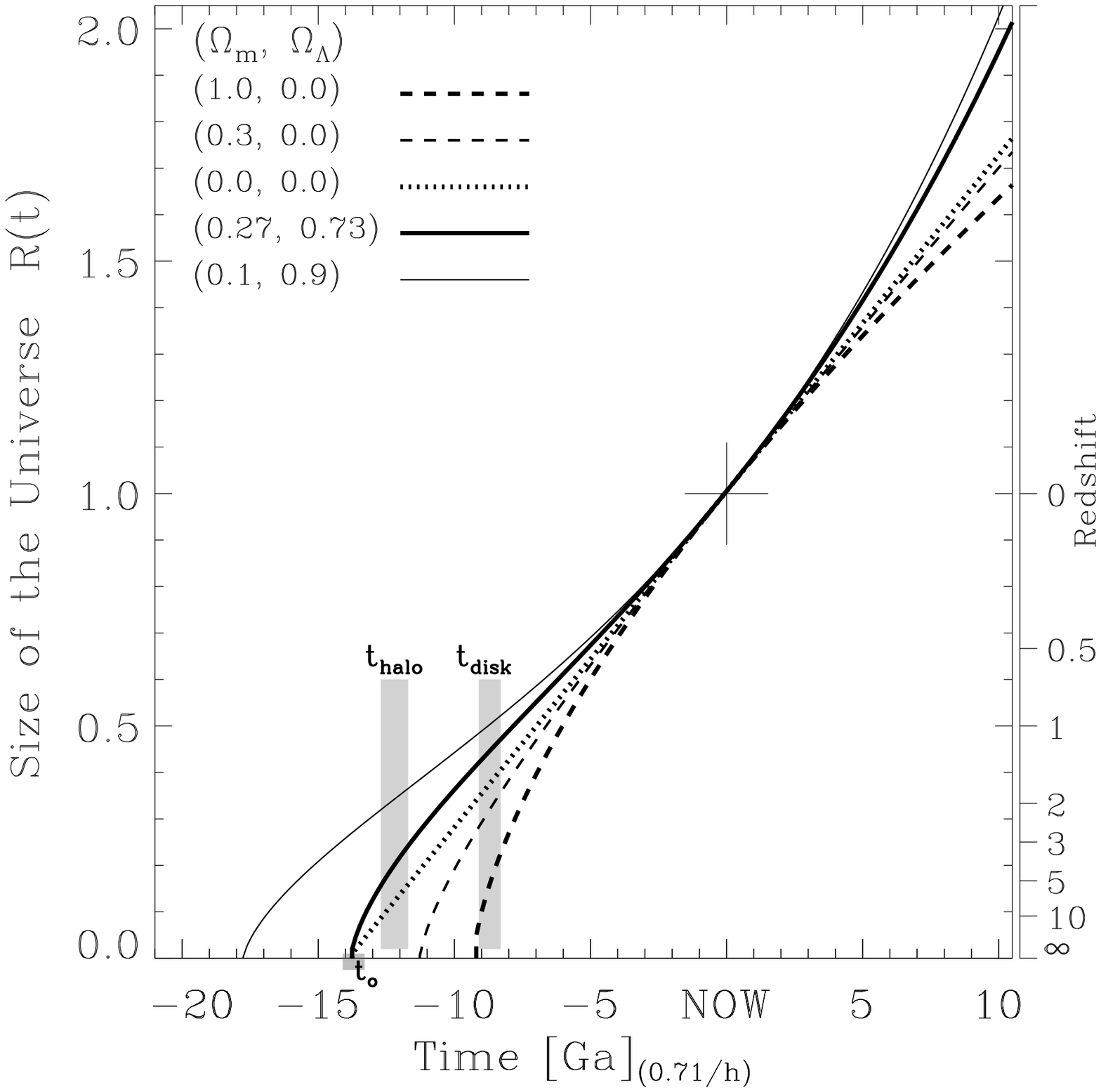} % postscript image file name
\caption{  {\bf  Size and Destiny of the Universe.}   %\\
This plot shows the size of the Universe, in units of its current size, as a function of time.
The age of the five models can be read from the x axis as the time between `NOW'
and the intersection of the model with the x axis. 
%The main age result from Lineweaver 1999a,$\t = 13.4 \pm 1.6$ Gyr, 
%is labeled $\t$ and is shaded grey on the x axis.
Models containing $\oll$ curve upward ($\ddot{R} > 0$) and are currently accelerating.
The empty universe has $\ddot{R} = 0$ (dotted line) and is `coasting'. 
The expansion of matter-dominated universes is slowing down ($\ddot{R} < 0$).
The $(\oll, \om) \approx (0.27, 0.73)$ model is favored by the data.
Over the past few billion years and  on into the future, the rate of expansion
of this model increases. This acceleration means that we are in a period of slow 
inflation -- a new period of inflation is starting to grab
the Universe. Knowing the values of $h$, 
$\om$ and $\oll$ yields a precise relation between age, redshift and size of
the Universe allowing us to convert the ages of local objects 
(such as the disk and halo of our galaxy) into redshifts. We can then examine 
objects at those redshifts to see if disks are forming at a redshift of $\sim 1$ 
and halos are forming at $z \sim 4$.
This is an example of th tightening network of constraints produced by precision cosmology.
}
\label{fig:GRexpansion}
\end{figure}
%%%%%%%%%%%%%%%%%%%%%%%%%%%%%%%%%%%%

\subsection{Observational Constraints from the CMB}
\label{sec:constraints}

Our general relativistic description of the Universe can be divided into two parts, those parameters
like $\Omega_{i}$ and $H$ which describe the global properties of the model and those
parameters like $n_{s}$ and $A$ which describe the perturbations to the global properties and hence
describe the large scale structure (Table 1).

In the context of general relativity and the 
hot big bang model, cosmological parameters are the numbers that, when
inserted into the Friedmann equation, 
%(Eq. 1)  % \ref{eq:Friedmann}) 
best describe our particular observable universe.
These include Hubble's constant $H$ (or $h = H/100\; km\; s^{-1} Mpc^{-1}$),
the cosmological constant $\oll = \Lambda/3H^{2}$, geometry $\ok = -k/H^{2}R^{2}$, the density
of matter, $\om = \oc + \ob = \rho_{CDM}/\rho_{c} + \rho_{baryon}/\rho_{c}$ 
and the density of relativistic matter $\orel = \og  + \on$.
Estimates for these have been derived from hundreds of observations and analyses.
Various methods to extract cosmological parameters from cosmic microwave
background (CMB) and non-CMB 
observations are forming an ever-tightening network of interlocking constraints.
CMB observations now tightly constrain  $\ok$, while type Ia supernovae 
observations tightly constrain the deceleration parameter $q_{o}$. Since lines of constant $\ok$ and 
constant $q_{o}$ are nearly orthogonal in the $\om-\oll$ plane, 
combining these measurements optimally constrains our Universe 
to a small region of parameter space.

The upper limit on the energy density of neutrinos comes from the shape of the
small scale power spectrum. If neutrinos make a significant contribution
to the density, they suppress the growth of small scale structure
by free-streaming out of over-densities.
The CMB power spectrum is not sensitive to such small scale power or its suppression,
and is not a good way to constrain $\on$. And yet the best limits on $\on$ come from the WMAP normalization of 
the CMB power spectrum used to normalize the power spectrum of galaxies from
the 2dF redshift survey (Bennett et al. 2003).

The parameters in Table 1 are not independent of each other. 
For example, the age of the Universe,  $t_{o} = h^{-1}f(\om, \oll)$.
If $\Omega_{m} = 1$ as had been assumed by most theorists until about 1998, then the age of the Universe
would be simple:
\beq
t_{o}(h) = \frac{2}{3}H_{o}^{-1} = 6.52 \; h^{-1} Gyr.
\eeq
However, current best estimates of the matter and vacuum energy densities are $(\om,\oll) = (0.27, 0.73)$.
For such flat universes ($\Omega = \om  + \oll = 1$) we have (Carroll et al. 1992):
\beq
t_{o}(h,\om,\oll) = \frac{1}{\sqrt{\oll}}\left(\frac{1+\sqrt{\oll}}{\sqrt{\om}}\right) 6.52 \; h^{-1} Gyr.
\eeq
for  $t_{o}(h = 0.71, \om = 0.27, \oll = 0.73) = 13.7$ Gyr.

If the Universe is to make sense, independent
determinations of $\oll$, $\om$ and $h$ and the minimum age of the Universe
must be consistent with each other.
This is now the case (Lineweaver 1999).
Presumably we live in a universe 
which corresponds to a single point in 
multidimensional parameter space.
Estimates of $h$ from HST Cepheids
and the CMB overlap. Deuterium and CMB determinations of $\obh$
are consistent. Regions of the $\om - \oll$ plane favored by supernovae
and CMB overlap with each other and with other independent constraints 
(e.g. Lineweaver 1998).
The geometry of the Universe does not seem to 
be like the surface of a ball ($\ok < 0$) nor like a saddle ($\ok > 0$) 
but seems to be flat ($\ok \approx 0$) to the precision of our 
current observations.

There has been some speculation recently that the evidence for $\oll$ is really 
evidence for some form of stranger dark energy (dubbed `quintessence') that 
we have been incorrectly interpreting as $\oll$. 
The evidence so far indicates that the cosmological constant 
interpretation fits the data as well as or better than an explanation based 
on quintessence.
% more mysterious dark energy.
% (Perlmutter et al. 1999a, Garnavich et al. 1998, 
%Perlmutter et al. 1999b).

\begin{table}
\begin{center}
\caption{\label{cosparam} Cosmological parameters describing the best-fitting FRW model to the CMB power
spectrum and other non-CMB observables (cf. Bennett et al. 2003).}
\vspace{2pt}
%%%%%%%%%%%%%%%%%%%%%%%%%%%%%%%%%%%%%%%%%%%%%%  table modified from Bennett etal Basic Results paper   %%%%%%%%%%%
%\begin{deluxetable}{lcccc}
\begin{tabular}{lcc} \hline
\multicolumn{3}{|c|}{Composition of Universe$^{a}$}\\
%\caption{``Best'' Cosmological Parameters  \label{tbl-2}}\\
%Description            & Symbol   &  Value & $+$ uncertainty &  $-$ uncertainty\\
\hline
Total density          &\ensuremath{\Omega_{o}}    &    \ensuremath{1.02 \pm 0.02}\\ 
Vacuum energy density    &\ensuremath{\Omega_\Lambda}  &    \ensuremath{0.73 \pm 0.04}\\
%Matter density         &\ensuremath{\Omega_m}        &   \ensuremath{0.27 \pm 0.04}\\
Cold Dark Matter density &\ensuremath{\Omega_{CDM}}   &   \ensuremath{0.23 \pm 0.04}\\
Baryon density         &\ensuremath{\Omega_b}        &   \ensuremath{0.044 \pm 0.004}\\ 
Neutrino density &$\Omega_\nu$                 &       $<0.0147$  95\% CL  \\
Photon density         &\ensuremath{\Omega_{\gamma}} &   \ensuremath{4.8 \pm 0.014 \times 10^{-5}}\\ 
\hline
\multicolumn{3}{|c|}{Fluctuations}\\
\hline
%Fluctuation amplitude in $8h^{-1}$ Mpc spheres       &\ensuremath{\sigma_8}     &\ensuremath{0.84 \pm 0.04}\\
Spectrum normalization$^{b}$      &$A$& \ensuremath{0.833^{+0.086}_{-0.083}}\\
Scalar spectral index$^{b}$ 
                       &\ensuremath{n_s}             &   \ensuremath{0.93 \pm 0.03}\\
Running index slope$^{b}$
                       &\ensuremath{dn_s/d\ln{k}}    &   \ensuremath{-0.031^{+0.016}_{-0.018}}\\
Tensor-to-scalar ratio$^{c}$                         &   \ensuremath{r=T/S}
                       &$<0.71$ 95\% CL\\
\hline
\multicolumn{3}{|c|}{Evolution}\\
\hline
Hubble constant        &\ensuremath{h}               &   \ensuremath{0.71^{+0.04}_{-0.03}}\\
Age of Universe (Gyr)  &\ensuremath{t_0}             &   \ensuremath{13.7 \pm 0.2}\\
Redshift of matter-energy equality &\ensuremath{z_{eq}}     &\ensuremath{3233^{+194}_{-210}}\\
Decoupling Redshift &\ensuremath{z_{dec}}         &   \ensuremath{1089 \pm 1}\\
Decoupling epoch (kyr) &\ensuremath{t_{dec}}        &   \ensuremath{379^{+8}_{-7}}\\
Decoupling Surface Thickness (FWHM) &\ensuremath{\Delta z_{dec}}     &\ensuremath{195 \pm 2}\\
Decoupling duration (kyr) &\ensuremath{\Delta t_{dec}}     &\ensuremath{118^{+3}_{-2}}\\
Reionization epoch (Myr, 95\% CL)) &\ensuremath{t_r}&   \ensuremath{180^{+220}_{-80}}\\
Reionization Redshift  (95\% CL) &\ensuremath{z_r} & \ensuremath{20^{+10}_{-9}}\\
Reionization optical depth &\ensuremath{\tau}     &\ensuremath{0.17 \pm 0.04}\\
\hline
\end{tabular}
\end{center}
$^{a}$ $\Omega_{i} = \rho_{i}/\rho_{c}$ where $\rho_{c} = 3H^{2}/8\pi G$\\
$^{b}$ at a scale corresponding to wavenumber $k_0=0.05$ Mpc$^{-1}$\\
$^{c}$ at a scale corresponding to wavenumber $k_0=0.002$ Mpc$^{-1}$
\end{table}
%%%%%%%%%%%%%%%%%%%%%%%%%%%%%%%%%%%%%%%%%%%%%%%%%%%%%%%%%%%%%%%%%%%%%%%%%%%%%%%%%%%%%%%%%

\subsection{Background and the Bumps on it and the Evolution of those Bumps}

Equation \ref{eq:fried1} is our hot big bang description 
of the unperturbed Friedmann-Robertson-Walker universe.  
There are no bumps in it, no 
over-densities, no inhomogeneities, no anisotropies 
and no structure. 
The parameters in it are the background parameters.
It describes the evolution of a perfectly
homogeneous universe. 

However, bumps are important.
If there had been no bumps in the CMB thirteen billion 
years ago, no structure would exist today.
The density bumps seen as the hot and cold spots in the CMB map have 
grown into gravitationally enhanced light-emitting over-densities known 
as galaxies (Fig. \ref{fig:2df}). Their gravitational growth depends on the 
cosmological parameters -- 
much as tree growth depends on soil quality 
(see Efstathiou 1990 for the equations of evolution of the bumps).
We measure the evolution of the bumps and from them we infer the background.
Specifically, matching
the power spectrum of the CMB (the $C_{\ell}$'s which sample the 
$z \sim 1000$ universe) to the power spectrum of local galaxies (the $P(k)$ 
which sample the $z \sim 0$ universe) we can constrain 
cosmological parameters. The limit on $\on$ is an example.

\subsection{The End of Cosmology?}
When the WMAP results came out at the end of this school I was asked
``So is this the end of cosmology? We know all the cosmological parameters...what is there left to do?
To what precision does one really want to know the value of $\Omega_{m}$?''
In his talk, Brian Schmidt asked the rhetorical question: ``We know Hubble's parameter to about 10\%, is that
good enough?'' Well, now we know it to about 5\%. Is that good enough?
Obviously the more precision on any one parameter the better, but we are talking about constraining an entire
model of the universe defined by a network of parameters.
As we determine 5 parameters to less than 10\%, it enables us to turn a former upper limit on another parameter into
a detection. For example we still have only upper limits on the tensor to scalar ratio $r$ and 
this limits our ability to test inflation. We only have an upper limit on
the density of neutrinos $\Omega_{\nu}$ and this limits our ability  to go beyond the standard model
of particle physics. And we have only a tenuous detection of the running of the scalar spectral index
$dn/dln k \neq 0$, and this limits our ability to constrain inflaton potential model builders.

We still know next to nothing about $\oll \sim 0.7$, most of the Universe.
$\Lambda$CDM is an observational result that has yet to be 
theoretically confirmed. From a quantum field theoretic point of view
$\oll \sim 0.7$ presents a huge problem. It is a quantum term in
a classical equation. But the last time such a 
quantum term appeared in a classical equation, Hawking radiation 
was discovered.
A similar revelation may be in the offing.
The Friedmann equation will eventually be seen as a low energy approximation
to a more complete quantum model in much the same way that 
$\frac{1}{2} m v^{2}$ is a low energy approximation
to $pc$.

Inflation solves the origin of structure problem with quantum fluctuations,
and this is just the beginning of quantum contributions to cosmology.
Quantum cosmology is opening up many new doors.
Varying coupling constants are expected at high energy
(Wilczek 1999) and $c$ variation, $G$ variation,
$\alpha$  (fine structure constant) variation, and $\oll$ variation (quintessence)
are being discussed. 
We may be in an ekpyrotic universe or a cyclic one
(Steinhardt \& Turok 2002).
The topology of the Universe is also alluringly fundamental
(Levin 2002).
Just as we were getting precise estimates of the parameters
of classical cosmology,
whole new sets of quantum cosmological parameters are being proposed.
The next high profile goal of cosmology may be trying to figure out if we are living in a multiverse.
And what, pray tell, is the connection between inflation and dark matter?

\subsection{Tell me More}

For a well-written historical (non-mathematical) review of inflation see Guth (1997).
For a detailed mathematical description of inflation see Liddle and Lyth (2000).
For a concise mathematical summary of cosmology for graduate students see Wright (2003).
Three authoritative texts on cosmology that include inflation and the CMB are
`Cosmology' by P.~Coles and F.~Lucchin,
`Physical Cosmology' by P.~J.~E.~Peebles and
`Cosmological Physics' by J. Peacock.

\section*{Acknowledgments}
I thank Mathew Colless for inviting me to give these five lectures to such
an appreciative audience.
I thank John Ellis for useful discussions as we bushwhacked in the gloaming.
I thank Tamara Davis for Figs.~\ref{fig:triptych}, \ref{fig:fig1bot} \& \ref{fig:verywide}.
I thank Roberto dePropris for preparing Fig. \ref{fig:2df}.
I thank Louise Griffiths for producing Fig. \ref{fig:binnedcls} and Patrick Leung
for producing Figs. \ref{fig:BWcleaned512} \& \ref{fig:twomaps}.
The HEALPix package (G\'{o}rski, Hivon and Wandelt 1999) was used to prepare these maps.
I acknowledge a Research Fellowship from the Australian Research Council.

\end{document}